\documentclass[a4paper,10pt,dvipsnames]{article}
\usepackage[utf8]{inputenc}
\usepackage[english]{babel}
\usepackage{amsthm}
\usepackage{amsfonts}
\usepackage{amssymb}	
\usepackage{amsmath}
\allowdisplaybreaks
\usepackage{mathtools}
\usepackage{slashed}
\usepackage{enumerate}
\usepackage{graphicx}
\usepackage{systeme}
\usepackage{dsfont}
\usepackage{relsize}
\usepackage[margin=0.90in]{geometry}
\usepackage{float}
\usepackage{tikz}
\usepackage[labelformat=simple]{subcaption}

\usepackage{graphicx}
\usepackage{bmpsize}
\usepackage{epstopdf}
\usepackage{bbm}
\usepackage{etoolbox}
\usepackage{xcolor}
\usepackage{titling}
\usepackage{authblk}
\newcommand*\samethanks[1][\value{footnote}]{\footnotemark[#1]}
\usepackage{blindtext,graphicx}
\usepackage[absolute]{textpos}
\usepackage{physics}
\usepackage[hang,flushmargin]{footmisc}

\usepackage{xfrac}
\usepackage{nicefrac}
\usepackage{esvect}
\usepackage{cases}
\usepackage{empheq}
\usepackage{stmaryrd}
\usepackage{cancel}
\usepackage[linguistics]{forest}
\usepackage{url}
\usepackage[font=small]{caption}
\usepackage{array}
\usepackage[giveninits=true,sortcites=true,date=year,maxbibnames=99,doi=false,isbn=false,url=false,eprint=false]{biblatex}
\renewbibmacro{in:}{}
\DeclareFieldFormat{pages}{#1}
\AtEveryBibitem{%
  \clearlist{language}%
}
\usepackage{csquotes}

\usepackage{setspace}

\title{Towards the conception of complex engineering meta-structures: relaxed-micromorphic modelling of low-frequency mechanical diodes\!\! /high-frequency screens}
\author{
Gianluca Rizzi\thanks{GEOMAS, INSA-Lyon, Université de Lyon, 20 avenue Albert Einstein, 69621, Villeurbanne Cedex, France}
,\quad
Domenico Tallarico\thanks{Laboratory for Acoustics/ Noise Control, Swiss Federal Laboratories for Materials Science and Technology (EMPA), \\ \indent Überlandstrasse 129, 8600, Dübendorf, Switzerland}
,\quad
Patrizio Neff\thanks{Head of Chair for Nonlinear Analysis and Modelling, Fakultät für Mathematik, Universität Duisburg-Essen, \\ \indent Thea-Leymann-Straße 9, 45127 Essen, Germany}
,\quad
Angela Madeo\samethanks[1]
}

\thanksmarkseries{arabic}
\date{\today}

\makeindex

\begin{document}
\maketitle
\begin{abstract}
\begin{small}
In this paper we show that an enriched continuum model of the micromorphic type (Relaxed Micromorphic Model) can be used to model metamaterials' response in view of their use for meta-structural design.
We focus on the fact that the reduced model's structure, coupled with the introduction of well-posed interface conditions, allows us to easily test different combinations of metamaterials' and classical-materials bricks, so that we can eventually end-up with the conception of a meta-structure acting as a mechanical diode for low/medium frequencies and as a total screen for higher frequencies.
Thanks to the reduced model's structure, we are also able to optimize this meta-structure so that the diode-behaviour is enhanced for both ``pressure" and ``shear" incident waves and for all possible angles of incidence.
\end{small}
\end{abstract}

\vspace{2mm}
\section{Introduction}
\label{sec:intro}
The last decade has seen the birth of a true research outburst on so-called mechanical metamaterials which are able to show exotic mechanical properties both in the static and dynamic regime.
Theoretical, experimental and numerical studies have flourished all around the world providing new insights in the domain of materials’ properties manipulation which, only few years ago, was thought far from being prone to possible ground-breaking evolutions. We are today seeing the conception and subsequent realization of new materials which, simply thanks to their internal architecture, go beyond the materials’ properties that we are used to know and which, for this reason, are called metamaterials.
It is thus possible today to see 3D-printed pyramids connected by hinges giving rise to a block that is stiff like a brick on one side but compliant like a sponge on the other \cite{bilal2017intrinsically}, ``unfeelability'' cloaks hiding to the touch objects put below them \cite{kadic2014pentamode,milton1995elasticity}, plastic cubes made out of smaller plastic cubes giving rise to bizarre deformations when squeezed \cite{coulais2016combinatorial}, or even metamaterials exploiting microstructural instabilities to change their mechanical response depending on the level of externally applied load \cite{kochmann2017exploiting}.
When considering the dynamical behaviour of mechanical metamaterials, things become even more impressive, given the unorthodox responses that such metamaterials can provide when coming in contact with elastic waves \cite{deymier2013acoustic,hussein2014dynamics}.
It is today possible to find researchers designing metamaterials exhibiting band-gaps \cite{celli2019bandgap,bilal2018architected,liu2000locally,wang2014harnessing}, cloaking \cite{buckmann2015mechanical,misseroni2016cymatics}, focusing \cite{cummer2016controlling,guenneau2007acoustic}, channelling \cite{kaina2017slow,tallarico2017edge,bordiga2019prestress}, negative refraction \cite{willis2016negative,bordiga2019prestress}, etc., as soon as they interact with mechanical waves.

Notwithstanding this success on unveiling always new metamaterials’ performances, their application is still drastically limited. This is because we lack models that can predict how metamaterials’ properties are modified when different metamaterials’ bricks are combined together.
It is clear that we need to know what happens when these bricks are combined and which proprieties are enhanced, if we want to use metamaterials to build realistic devices for wave manipulation and control.
The step from the conception of new metamaterials to their use in meta-structural design cannot be realistically accomplished using direct finite element simulations accounting for all the details of the underlying microstructures. This would lead to unaffordable computational costs already for structures counting few dozens of unit cells.
The awareness of this limitation triggered all the recent advances on dynamical homogenization methods \cite{chen2001dispersive,willis2009exact,craster2010high,willis2011effective,willis2012construction,boutin2014large,sridhar2018general}.
Such methods share the idea that a periodic infinite-size metamaterial can be replaced by a homogenized continuum, mimicking its response without accounting for all the microstructures’ details.
This leads to an important simplification of metamaterials’ description at the macroscopic scale.
However, as soon as finite-size metamaterials are concerned, it is often difficult to set up well-posed boundary value problems that are representative of the static/dynamic behaviour of large classes of finite-size metamaterials.

In the recent past some of the authors suggested that a suitable framework to deal with finite-size metamaterials modelling at the macroscopic scale is micromorphic continuum mechanics.

Micromorphic models extend classical Cauchy elasticity by introducing additional degrees of freedom with respect to the macroscopic displacement field, typically through a non-symmetric, second order tensor (micro-distortion tensor P) \cite{mindlin1964micro,eringen1968mechanics,madeo2016first,neff2014unifying}. These additional degrees of freedom allow to account for micro-motions that are ``a priori'' independent of the macroscopic displacement (they can be, e.g., interpreted as motions of the microstructure inside a metamaterial’s unit-cell). Such extension of the kinematic framework, coupled with the introduction of suitable constitutive laws, results in a generalization of the governing equations of the considered continuum with respect to Cauchy elasticity. Such generalized framework allows to describe complex dynamic metamaterials behaviors via the definition of additional micro and macro elasticity and micro-inertia tensors that i) are frequency independent and ii) encode the static and dynamic anisotropy of the considered metamaterial. In this generalized framework, the concept of effective mass density remains unvaried with respect to its more classical definition. In the literature, important efforts have been made to describe complex dynamic metamaterial behaviors via the introduction of an anisotropic, frequency-dependent mass density \cite{milton2007modifications,willis1985nonlocal} that can, in some cases, even become negative \cite{avila2005bandes,liu2005analytic,mei2006effective}. The need of giving an anisotropic, frequency-dependent structure to the mass density is directly related to the fact that the metamaterials’ governing equations are enforced to follow classical Cauchy elasticity. In other words, at each frequency, the metamaterial is locally regarded as a Cauchy medium that has a different density (and, eventually, different elastic properties), so that the typical metamaterial’s dispersive behavior can be recovered. On the contrary, using a micromorphic framework, all the elastic and inertia parameters (as well as the mass density) remain frequency-independent thus allowing a metamaterial’s characterization through the calibration of a limited number of frequency-independent parameters. In this sense, we can say that micromorphic metamaterials’ characterization follows more closely the classical concept of materials’ characterization through the introduction of few constant parameters (e.g., density, Young modulus and Poisson-ratio when considering an isotropic, homogeneous material).

In \cite{madeo2015wave} the \textit{relaxed micromorphic model} has been introduced and used to describe the dynamical behaviour of band-gap metamaterials.
Optimized relaxed-micromorphic constitutive laws \cite{neff2014unifying} were then proposed to characterize realistic 1D metamaterials \cite{madeo2016first,madeo2018relaxed} and semi-analytical solutions for the frequency-dependent scattering of a relaxed micromorphic half-plane were found \cite{aivaliotis2019microstructure}, thus providing deeper understanding on the fundamental problem of establishing well-posed boundary conditions in micromorphic media.
The relaxed micromorphic model was also calibrated to describe the average behaviour of certain infinite-size 2D metamaterials \cite{madeo2018modeling,barbagallo2019relaxed}.
Finally, some of the authors investigated how boundary conditions should be introduced in micromorphic media to provide well-posed boundary value problems for 2D finite-size tetragonal metamaterials \cite{madeo2018relaxed,aivaliotis2019microstructure,aivaliotis2020frequency}.
Thanks to these preliminary works, it was established that the micromorphic modelling of finite-size metamaterials can indeed open new perspectives towards the conception and design of complex meta-structures that can control elastic waves and recover energy.
To demonstrate the effectiveness of the proposed micromorphic approach to design useful metamaterials’ devices we focus in this paper
on the conception of a meta-structure that acts as a mechanical diode at low/medium frequencies and as a screen at higher frequencies. A mechanical diode is a device that allows elastic wave transmission when the wave propagates in one direction, while preventing it when considering the opposite direction of propagation.

On the other hand, a screen can be seen as a device which entirely reflects elastic waves, independently of their direction of propagation.

In the literature, scientists are mainly focusing on trying to engineer the internal metamaterial’s architecture following non-homogeneous patterns (see e.g. \cite{gliozzi2019proof,grinberg2018acoustic,baz2018active,bennett2019acoustic,fu2018high,li2019diode,popa2014non,wang2016broadband,wei2020nonreciprocal,maznev2013reciprocity,parnell2013antiplane}) or non-elastic constitutive laws \cite{nadkarni2016unidirectional} so as to create diodes for screens.

We show in the present paper that the desired properties can indeed be achieved by suitably embedding a metamaterial's slab between two homogeneous elastic half-spaces with different material properties.
Specifically, we suggest that the fact of combining a given metamaterial with classical homogeneous materials can drastically enhance the metamaterial's original properties.

In particular, we start with the observation that imposing a suitable difference of stiffness between the two external Cauchy media may trigger a low-frequency diode behaviour.
Based on this simple observation remark and exploiting the computational performances of the relaxed micromorphic model, we are able to explore different structure's configurations, finally ending up in a prototype of low-frequency diode/high frequency screen.
This configuration is then optimized to show the desired behaviour for large intervals of angles of incidence for both pressure and shear waves.

This opens the way to the conception of meta-structures with many extra possible functionalities with respect to those that could be possible by simply considering the metamaterial alone.

In this paper we present simple meta-structures that can act as protection tools if placed around an object (complete reflection if the incident wave comes from the exterior of the domain that we want to preserve) while they allow transmission when the incident wave comes from the protected environment itself.
The same structures can screen waves in both directions when suitably increasing the wave's frequency.

It is evident that the possibilities of realistic metamaterials' use in real meta-structures is multiplied by the fact of combining them with classical materials (and, in general, with other metamaterials) so as to create complex meta-structures that control elastic wave propagation.

The present paper opens new perspective for the use of metamaterials in meta-structural design, even if the proposed relaxed micromorphic model will need considerable extension to provide broadband quantitative accuracy.

\subsection{Notation}
\label{sec:notation}
Vectors will be denoted with a lower case letter, second order tensors with upper case letters, and fourth order tensor with a blackboard bold upper case letter.
A simple contraction between tensors is denoted by $\cdot$, while the scalar product is denoted by $\left\langle \cdot, \cdot \right\rangle$.
The Einstein contraction is implied throughout this text unless otherwise specified.
The Frobenius tensor norm is $\lVert X \rVert^2 = \left\langle X, X\right\rangle$.
The identity tensor on $\mathbb{R}^{3\times 3}$ will be denoted by $\mathbbm{1}$ with $\mbox{tr}(X)=\left\langle X, \mathbbm{1} \right\rangle$.
We denote by $B_L$ a bounded domain in $\mathbb{R}^3$, by $\delta B_L$ its regular boundary, and by $\Sigma$ any material surface embedded in $B_L$.
The outward unit normal to $\delta  B_L$ and to a surface $\Sigma$ will be denoted by $\nu$.
Given a field $a$, we define its jump through the surface $\Sigma$ as
\begin{equation}
\llbracket a \rrbracket = a^+ - a^-, \qquad \text{with}
\qquad
a^{-} \coloneqq \lim_{\substack{x \in B_L^{-}\setminus \Sigma \\ x \to \Sigma}} a ,\qquad \text{and}
\qquad
a^{+} \coloneqq \lim_{\substack{x \in B_L^{+}\setminus \Sigma \\ x \to \Sigma}} a,
\end{equation} 
where $B_L^{-}, B_L^{+}$ are the two sub-domains which result from splitting $B_L$ by the surface $\Sigma$.

Classical gradient $\nabla$ and divergence $\mbox{Div}$ operators are used throughout the paper.
The subscript $,j$ implies the derivation with respect to the $j-$th component of the space variable, while the subscript $,t$ only denotes derivation with respect to time.\footnote{
	Being reserved to the time variable, the index \textit{t} is treated separately and does not comply with Einstein notation.
}

The classical macroscopic displacement field is denoted by $u(x,t)\in \mathbb{R}^3$, with $x\in B_L,\, t\in [0,t_0]$.
In the relaxed micromorphic model, extra kinematic degrees of freedom are added through the introduction of the non-symmetric micro-distortion $P(x,t) \in \mathbb{R}^{3\times 3}$, with $x\in B_L,\, t\in [0,t_0]$.
\section{Equilibrium equations, constitutive relations, and energy flux}
\label{sec1}
In this section, we present a summary of the governing equations and energy conservation law describing the macroscopic mechanical behaviour of both Cauchy and relaxed micromorphic media.
The classical Cauchy setting will be used to model the response of homogeneous materials, while the relaxed micromorphic model will be adopted to describe the metamaterial's response.
\subsection{Isotropic Cauchy continuum}
\label{subsec1}
The equilibrium equations for the classical isotropic Cauchy continuum are
\begin{equation}
\rho \, u_{,tt} = \mbox{Div}\left[\sigma\right] \, ,
\qquad
\mbox{with}
\qquad
\sigma = 2\mu \, \mbox{sym}\nabla u + \lambda \, \mbox{tr}\left(\mbox{sym} \nabla u\right) \mathbbm{1} \, ,
\label{eq:equiCau}
\end{equation}
where $\sigma$ is the Cauchy stress tensor, $\lambda$ and $\mu$ are the Lamé parameters and $\mbox{sym}\nabla u$ is the symmetric strain tensor.
When dissipative phenomena can be neglected, the following flux equation must hold
\begin{equation}
E_{,t} + \mbox{Div} H=0 \, ,
\label{eq:EnergyConservation}
\end{equation}
where $E$ is the total energy of the system and $H$ is the energy flux vector, whose explicit expression is given by (see e.g. \cite{aivaliotis2019microstructure} for a detailed derivation) 
\begin{equation}
H = -\sigma \cdot u_{,t} \, .
\label{eq:Cauchyflux}
\end{equation}

The reflection and transmission coefficients are defined as
\begin{align}
    \mathcal{R}=\frac{J^r}{J^i} \, ,
    \qquad\qquad
    \mathcal{T}=\frac{J^t}{J^i} \, ,
    \label{eq:ref_trans_coeff}
\end{align}
where 
\begin{align}
    J^i = \frac{1}{T} \int_{0}^{T} H^{i}(x,t) \, dt \, ,
    \qquad\qquad
    J^r = \frac{1}{T} \int_{0}^{T} H^{r}(x,t) \, dt \, ,
    \qquad\qquad
    J^t = \frac{1}{T} \int_{0}^{T} H^{t}(x,t) \, dt \, ,
    \label{eq:int_diff_J}
\end{align}
with the superscripts $i,r,t$ representing the incident, the reflected, and the transmitted wave contributions respectively, and due to the conservativeness of the framework in which we are working, we have that ${\mathcal{R} + \mathcal{T} = 1}$.
When considering a finite-size metamaterial's slab of the type shown in Fig.~\ref{fig:slab_mic} embedded between two Cauchy continua, it is sufficient to use eqs.(\ref{eq:ref_trans_coeff}) on the Cauchy sides to compute the energy which is reflected or transmitted across the slab. The fact that the considered slab is made of a micromorphic medium is accounted for through the use of the appropriate interface conditions eq.(\ref{eq:jumpdisplslab}) and eq.(\ref{eq:jumptractionslab}).
\subsection{Relaxed micromorphic continuum}
\label{subsec2}
The expression of the kinetic energy density for the relaxed micromorphic model is \cite{d2019effective,romano2016micromorphic}
\begin{equation}
\begin{array}{ll}
J \left(u_{,t},\nabla u_{,t}, P_{,t}\right) =& 
\dfrac{1}{2}\rho \, \langle u_{,t},u_{,t} \rangle + 
\dfrac{1}{2} \langle \mathbb{J}_{\rm micro}  \, \mbox{sym} \, P_{,t}, \mbox{sym} \, P_{,t} \rangle 
+ \dfrac{1}{2} \langle \mathbb{J}_{c} \, \mbox{skew} \, P_{,t}, \mbox{skew} \, P_{,t} \rangle \\[3mm]
& + \dfrac{1}{2} \langle \mathbb{T}_{e} \, \mbox{sym}\nabla u_{,t}, \mbox{sym}\nabla u_{,t} \rangle
  + \dfrac{1}{2} \langle \mathbb{T}_{c} \, \mbox{skew}\nabla u_{,t}, \mbox{skew}\nabla u_{,t} \rangle \, ,
\end{array}
\label{eq:kinEneMic}
\end{equation}
where $u$ is the macroscopic displacement field, $P \in \mathbb{R}^{3\times3}$ is the non-symmetric micro-distortion tensor, $\rho$ is the macroscopic apparent density, and $\mathbb{J}_{\rm micro}$, $\mathbb{J}_{c}$, $\mathbb{T}_{e}$, $\mathbb{T}_{c}$ are 4th order micro-inertia tensors whose form will be specified in the following subsection.

The expression of the strain energy density without curvature effects ($L_c=0$) is \cite{d2019effective,romano2016micromorphic}\footnote{
The presence of curvature terms is essential to catch size-effects in the static regime that are not the target of the present paper.
}
\begin{equation}
\begin{array}{ll}
W \left(\nabla u, P\right) =& 
  \dfrac{1}{2} \langle \mathbb{C}_{e} \, \mbox{sym}\left(\nabla u -  \, P \right), \mbox{sym}\left(\nabla u -  \, P \right) \rangle
+ \dfrac{1}{2} \langle \mathbb{C}_{\rm micro} \, \mbox{sym}  \, P,\mbox{sym}  \, P \rangle\\[3mm]
& 
+ \dfrac{1}{2} \langle \mathbb{C}_{c} \, \mbox{skew}\left(\nabla u -  \, P \right), \mbox{skew}\left(\nabla u -  \, P \right) \rangle
\, ,
\end{array}
\label{eq:strainEneMic}
\end{equation}
where $\mathbb{C}_{e}$, $\mathbb{C}_{\rm micro}$, and $\mathbb{C}_{c}$ are 4th order tensors whose characteristic will be given in Sec.~\ref{sec:plane_stain_symmetry}.
The minimization with respect to $P$ and $\nabla u$ of the action functional
\begin{equation}
\mathcal{A} = \int_{0}^{t_{0}} \int_{B_{L}} \left(J-W\right)dXdt
\label{eq:actionFuncMic}
\end{equation}
gives the following two sets of equilibrium equations
\begin{equation}
\rho \, u_{,tt} - \mbox{Div}\left(\widehat{\sigma}_{,tt}\right) = \mbox{Div}\left(\widetilde{\sigma}\right) \, ,
\qquad
\left( \mathbb{J}_{\rm micro} + \mathbb{J}_{c} \right) \, P_{,tt} = \widetilde{\sigma} - s \, ,
\label{eq:equiMic}
\end{equation}
where 
\begin{equation}
\begin{array}{cc}
\widehat{\sigma} \coloneqq \mathbb{T}_{e}~\mbox{sym} \nabla u + \mathbb{T}_{c}~\mbox{skew} \nabla u \, ,
\qquad
s \coloneqq \mathbb{C}_{\rm micro}~\mbox{sym}  \, P \, ,
\\[3mm]
\widetilde{\sigma} \coloneqq \mathbb{C}_{e}~\mbox{sym}\left(\nabla u -  \, P \right) + \mathbb{C}_{c}~\mbox{skew}\left(\nabla u -  \, P \right) \, .
\end{array}
\label{eq:sigMic}
\end{equation}

The flux equation for the relaxed micromorphic continuum is formally the same as eq.\eqref{eq:EnergyConservation}, but $H$ has now the following expression (see \cite{aivaliotis2020frequency} for more details):
\begin{equation}
H = -\left(\widetilde{\sigma} + \widehat{\sigma}_{,tt}\right)^T\cdot u_{,t} \, .
\label{eq:fluxAniso}
\end{equation}
The definitions of the reflection ($\mathcal{R}$) and transmission ($\mathcal{T}$) coefficients in eq. (\ref{eq:ref_trans_coeff})-(\ref{eq:int_diff_J}) still holds being careful to use the expression of the flux $H$ given in eq. (\ref{eq:fluxAniso}).

The structure for the relaxed micromorphic kinetic energy density (\ref{eq:kinEneMic}) and strain energy density (\ref{eq:strainEneMic}) has been chosen so as to minimize the number of introduced material parameters with respect to more classical micromorphic models \cite{mindlin1964micro,eringen1968mechanics} that may count more than 800 parameters in the general anisotropic case.
\subsection{Particularization of the relaxed micromorphic model to plane strain and tetragonal symmetry}
\label{sec:plane_stain_symmetry}
We now focus on finding solutions in a plane strain framework.
This means that we constrain the displacement field $u$ and the micro-distortion $P$ to depend only on the first two components $x_1$ and $x_2$ of the space variable $x$:

\begin{equation}
u (x_1,x_2) = 
\begin{pmatrix}
u_1(x_1,x_2)\\
u_2(x_1,x_2)\\
0\\
\end{pmatrix} \, ,
\quad
P (x_1,x_2) =  
\begin{pmatrix}
P_{11}(x_1,x_2)&P_{12}(x_1,x_2)&0\\
P_{21}(x_1,x_2)&P_{22}(x_1,x_2)&0\\
0&0&0\\
\end{pmatrix} \, .
\label{eq:plane_strain_u_u}
\end{equation}
The plane-strain hypothesis on the displacement field is retained also for the Cauchy media.

It is well known that the condition for positive definiteness of the strain energy density in the case of plane-strain isotropic Cauchy material reads:
\begin{align}
    \lambda + \mu > 0 \, , 
    \qquad\qquad
    \mu > 0 
\end{align}
where here $\widehat{\kappa} = \lambda + \mu$ is the plane strain bulk modulus.
Similarly, the positive definiteness of the strain energy density given in eq.(\ref{eq:strainEneMic}) for the tetragonal symmetry case in plane strain requires that
\begin{align}
    \lambda_{e}+ \mu_{e} > 0 \, , 
    \qquad
    \mu_{e} > 0  \, ,
    \qquad
    \mu_{e}^{*} > 0  \, ,
    \qquad
    \lambda_{\rm micro}+ \mu_{\rm micro} > 0  \, , 
    \qquad
    \mu_{\rm micro} > 0  \, , 
    \qquad
    \mu_{\rm micro}^{*} > 0 \, .
    \label{eq:pos_def_rmm}
\end{align}

Given the metamaterial targeted in this paper (see Fig.~\ref{fig:fig_tab_unit_cell_0}), we particularise the equilibrium equations to the tetragonal case.
This means that the micro inertia and the elastic tensors appearing in eq.(\ref{eq:kinEneMic})-(\ref{eq:strainEneMic}) can be represented in the Voigt form as

\begin{equation}
\begin{array}{rlrl}
	\mathbb{J}_{\rm micro} &=\rho 
	\begin{pmatrix}
	L^2_{3} + 2L^{2}_{1} & L^{2}_{3}              & \dots 			& \bullet \\ 
	L^{2}_{3}            & L^{2}_{3} + 2L^{2}_{1} & \dots 			& \bullet \\ 
	\vdots               & \vdots                 & \ddots 			& \bullet \\ 
	\bullet              & \bullet                & \bullet 		& L^{*^2}_{1} \\ 
	\end{pmatrix} \, ,
	\quad
	&\mathbb{J}_{c} &= \rho
	\begin{pmatrix}
	\bullet & 			& \bullet\\ 
	& \ddots 	& \vdots\\ 
	\bullet & \dots & 4L^{2}_{2}
	\end{pmatrix} \, ,
	\\[1cm]
	\mathbb{T}_{e} &= \rho
	\begin{pmatrix}
	\overline{L}^2_{3} + 2\overline{L}^2_{1}	& \overline{L}^2_{3}        			   	& \dots		& \bullet\\ 
	\overline{L}^2_{3}         				    & \overline{L}^2_{3} + 2\overline{L}^2_{1} 	& \dots		& \bullet\\ 
	\vdots                    					& \vdots                 				   	& \ddots 	&		 \\ 
	\bullet                   					& \bullet									& 	 		& \overline{L}^{*^2}_{1}
	\end{pmatrix} \, ,
	\quad
	&\mathbb{T}_{c} &= \rho
	\begin{pmatrix}
	\bullet & 			& \bullet\\ 
	& \ddots 	& \vdots\\ 
	\bullet & \dots & 4\overline{L}^2_{2}
	\end{pmatrix} \, ,
	\end{array}
\label{eq:micro_ine_2}
\end{equation}
\begin{equation}
    \begin{array}{rl}
    \mathbb{C}_{e} &= 
    \begin{pmatrix}
    \lambda_{e} + 2\mu_{e}	& \lambda_{e}				& \dots		& \bullet\\ 
    \lambda_{e}				& \lambda_{e} + 2\mu_{e}	& \dots		& \bullet\\ 
    \vdots					& \vdots					& \ddots	& 		 \\ 
    \bullet					& \bullet					& 			& \mu_{e}^{*}\\ 
    \end{pmatrix} \, ,
    \quad
    \mathbb{C}_{c} = 
    \begin{pmatrix}
    \bullet & 			& \bullet\\ 
    		& \ddots 	& \vdots\\ 
    \bullet & \dots		& 4\mu_{c}
    \end{pmatrix} \, ,
    \\[1cm]
    \mathbb{C}_{\rm micro} &= 
    \begin{pmatrix}
    \lambda_{\rm micro} + 2\mu_{\rm micro}	& \lambda_{\rm micro}				& \dots		& \bullet\\ 
    \lambda_{\rm micro}				& \lambda_{\rm micro} + 2\mu_{\rm micro}	& \dots		& \bullet\\ 
    \vdots					& \vdots					& \ddots	&  \\ 
    \bullet					& \bullet					& 			& \mu_{\rm micro}^{*}\\ 
    \end{pmatrix} \, ,
    \end{array}
\label{eq:micro_ine_1}
\end{equation}

where only the coefficients involved in a plane strain problem are reported (the dots represent components acting on out-of plane variables and are not specified here).

In the definition (\ref{eq:micro_ine_2}) of the micro-inertia tensors appearing in the kinetic energy (\ref{eq:kinEneMic}), it is underlined the fact that they introduce dynamic internal lengths that can be directly related to the dispersion behaviour of the metamaterial at very small (in the limit vanishing) wavenumbers ($\mathbb{J}_{\rm micro}$,$\mathbb{J}_{c}$), as well as at very large (in the limit infinite) wavenumbers ($\mathbb{T}_{e}$, $\mathbb{T}_{c}$).
\section{Interface conditions for a finite-size relaxed micromorphic slab embedded between two different Cauchy half-spaces}
\label{sec:BCs}
\begin{figure}[H]
	\centering
\begin{subfigure}[H]{0.45\textwidth}
	\includegraphics[width=\textwidth]{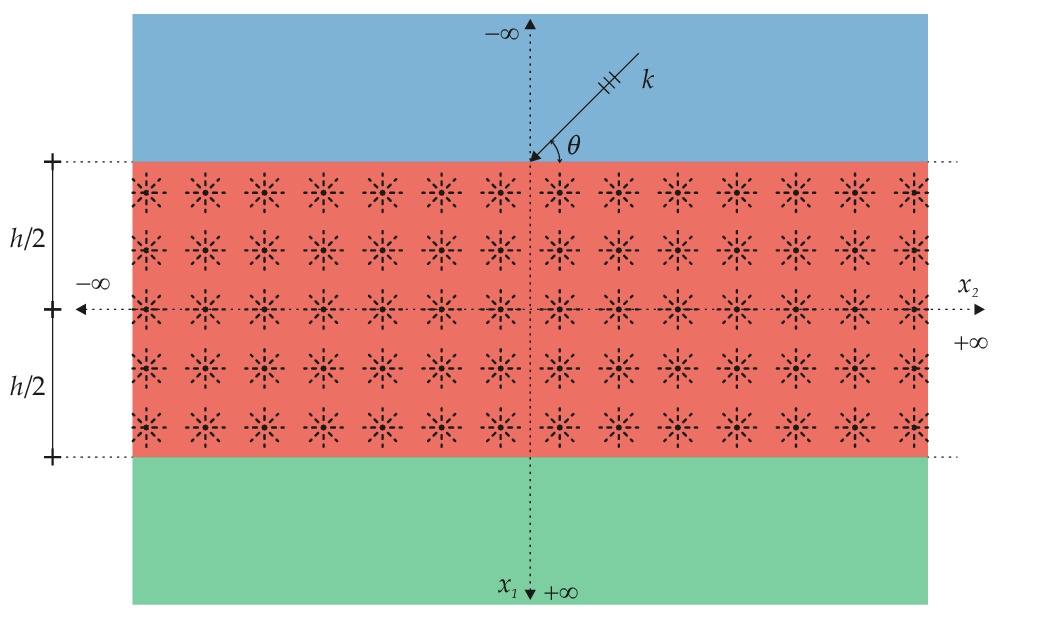}
	\caption{}
\end{subfigure}
\hfill
\begin{subfigure}[H]{0.45\textwidth}
	\includegraphics[width=\textwidth]{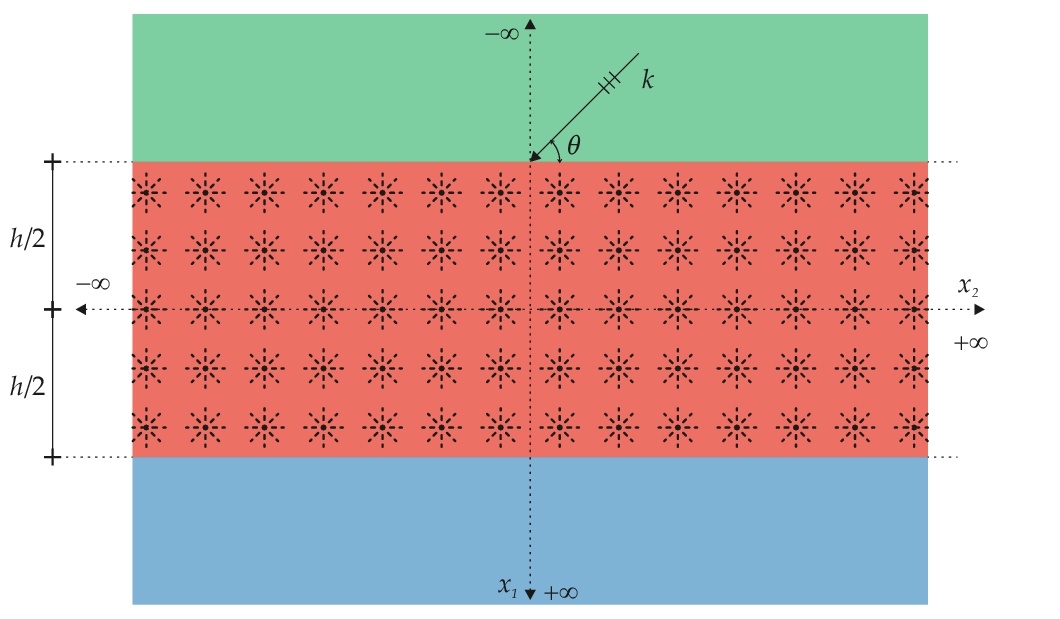}
	\caption{}
\end{subfigure}
	\caption{Schematic representation of a wave with wavenumber $k$ hitting at angle $\theta$ a relaxed micromorphic slab of thickness $h$ embedded between two isotropic Cauchy media with different stiffnesses.}
	\label{fig:slab_mic}
\end{figure}

A micromorphic slab of finite width $h$ is embedded between two different homogeneous Cauchy materials.
The material on the top is a classical linear elastic isotropic Cauchy medium, the material in the middle is an anisotropic relaxed micromorphic medium, while the material on the bottom is again a classical isotropic Cauchy medium with, \textit{a priori}, different stiffness with respect to the first homogeneous material (see Fig.~\ref{fig:slab_mic}). At the two interfaces these three materials are in perfect contact with each other.

At the interface between a classical Cauchy and a relaxed micromorphic material there are two conditions which can be imposed if the static characteristic length ($L_c$) is zero (our case here) \cite{madeo2016reflection,aivaliotis2019microstructure}: the continuity of displacement and continuity of generalized traction.

In the considered 2D case, there are then eight scalar conditions, four on each interface. The finite slab has width $h$ and we assume that the two interfaces are positioned at $x_1=-h/2$ and $x_1=h/2$, respectively (see Fig.~\ref{fig:slab_mic}). The continuity of displacement conditions to be satisfied at the two interfaces of the slab are
\begin{equation}\label{eq:jumpdisplslab}
u^{-}_c = u_m \quad \text{on} \quad x_1=-\frac{h}{2},
\qquad\qquad
u_m = u^{+}_c \quad \text{on} \quad x_1=\frac{h}{2},
\end{equation}
where $u^{-}_c$ and $u^{+}_c$ are the displacement of the ``minus'' ($x_1<0$) and ``plus''($x_1>0$) Cauchy half-space, respectively while $u_m$ is the displacement field in the relaxed micromorphic model slab.
The continuity of generalized traction reads
\begin{equation}\label{eq:jumptractionslab}
t^{-}_c = t_m \quad \text{on} \quad x_1=-\frac{h}{2},
\qquad\qquad
t_m = t^{+}_c \quad \text{on} \quad x_1=\frac{h}{2},
\end{equation}
where $t_{c}^{\pm} = \sigma^{\pm} \cdot \nu^{\pm}$ are classical Cauchy tractions, $t_{m} = \left(\widetilde{\sigma} + \widehat{\sigma}_{,tt}\right) \cdot \nu$ is the generalized traction in the relaxed micromorphic medium, with $\nu$ being the outward unit normal to the surface considered (see eq.(\ref{eq:sigMic}) for the definitions of generalized tractions).\footnote{
it is well known that the concept of traction results to be generalized in the framework of micromorphic continua with respect to the classical Cauchy traction.
This generalization is done through the generalization of the concept of stress that, in the particular case of the relaxed micromorphic material is given by the term inside the Div operator in eq.(\ref{eq:equiMic})$_{1}$.
}

Although the theoretical framework for the governing equations and interface conditions of the relaxed micromorphic material has been established in previous papers \cite{madeo2015band,madeo2015wave,madeo2016first}, the efficiency of the model to describe the refractive behaviour of finite-size meta-structures is only at its beginning \cite{aivaliotis2020frequency,rizzi2020exploring}.
In the present paper, we want to clearly establish that the proposed relaxed micromorphic model can be used for the conception of realistic meta-structure that control elastic waves by enabling the combination of homogeneous materials and metamaterials.
In particular, we will show that, thanks to the versatility of the relaxed micromorphic model, we are able to conceive a low-frequency diode/high-frequency screen.
Once the targeted meta-structures are designed via the relaxed micromorphic model, we check that they are effectively describing the behaviour of a realistic meta-structure by implementing the corresponding finite element simulation accounting for all microstructure's details.
The results obtained in this paper are a necessary step to proceed towards the use of the relaxed micromorphic model for more and more complex meta-structures that control elastic waves and recover energy.

\section{Relaxed micromorphic modelling of a 2D tetragonal metamaterial for acoustic control}
We briefly recall in this section how the relaxed micromorphic model can be used to model the broadband response of a tetragonal 2D metamaterial that was recently conceived for acoustic control \cite{rizzi2020exploring}.

\begin{figure}[H]
	\begin{subfigure}{0.49\textwidth}
		\centering
		\includegraphics[width=0.5\textwidth]{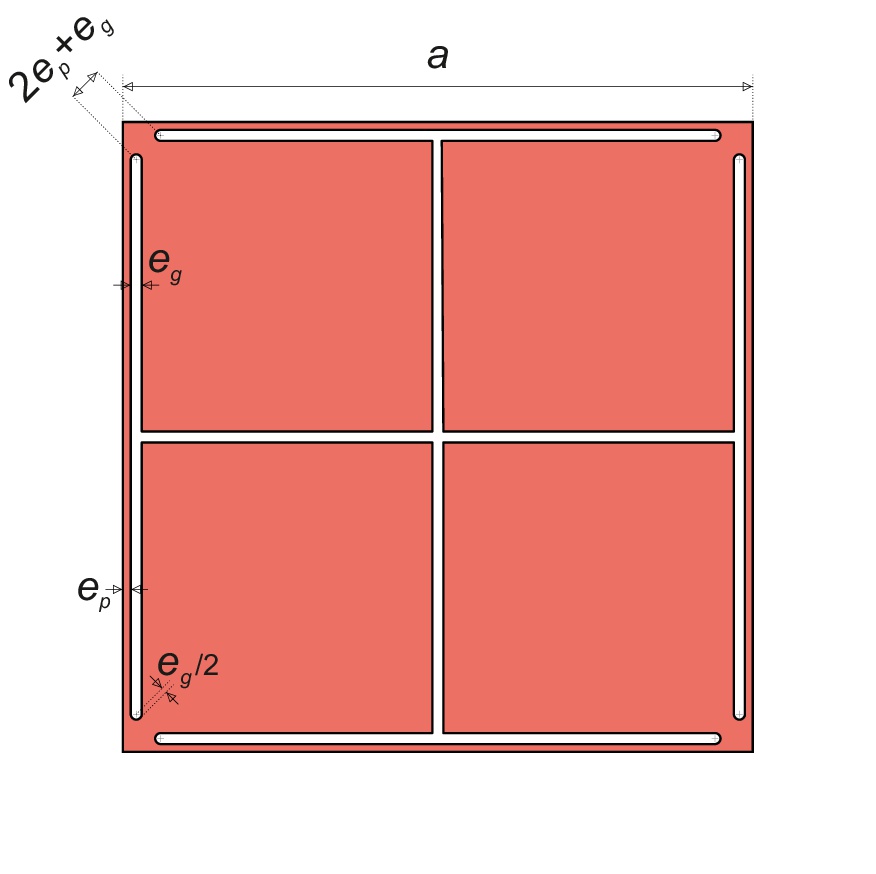}
		\caption{}
	\end{subfigure}
	\hspace*{-1.5cm}
	\begin{minipage}{0.49\textwidth}
		\begin{subfigure}{\textwidth}
			\centering
			\vspace{0.7cm}
			\begin{tabular}{cccccc}
				\hline
				$a$ & $e_{g}$ & $e_{p}$      \\ 
				$[$mm$]$ &  [mm]   &  [mm]   \\
				\hline
				20 & 0.35 & 0.25 \\ 
				\hline
				$\rho_{\rm Ti}$ & $\lambda_{\rm Ti}$ & $\mu_{\rm Ti}$          \\
				$\mbox{[kg/m}^3\mbox{]}$ &   [GPa]  &    [GPa] \\
				\hline
				4400 & 88.8 & 41.8\\ 
				\hline
				\vspace{0.2cm}
			\end{tabular}
			\caption{}
		\end{subfigure}
	\end{minipage}
	\caption{(a) unit cell whose periodic repetition in space gives rise to the metamaterial 1, or in short \textit{MM1}. (b) Table of the geometry and material properties of the unit cell: $\rho_{\rm Ti}$, $\lambda_{\rm Ti}$, and $\mu_{\rm Ti}$ stand for the density and the Lamé constants of titanium, respectively.}
	\label{fig:fig_tab_unit_cell_0}
\end{figure}

\begin{figure}[H]
	\begin{subfigure}[b]{0.45\linewidth}
		\includegraphics[width=\textwidth]{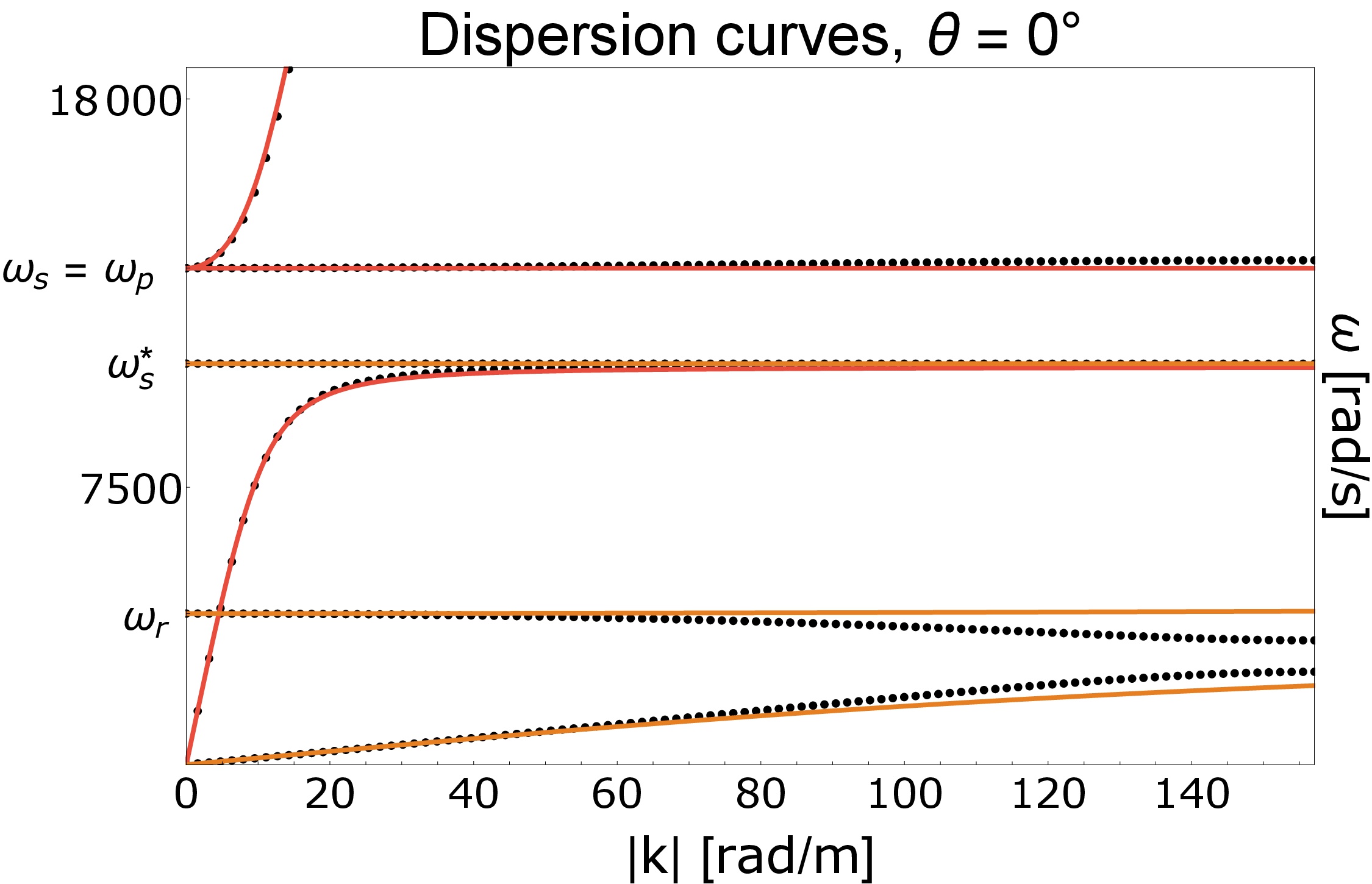}
		\caption{}
		\label{fig:disp_0_0}
	\end{subfigure}
	\hspace{0.5cm}
	\begin{subfigure}[b]{0.45\linewidth}
		\includegraphics[width=\textwidth]{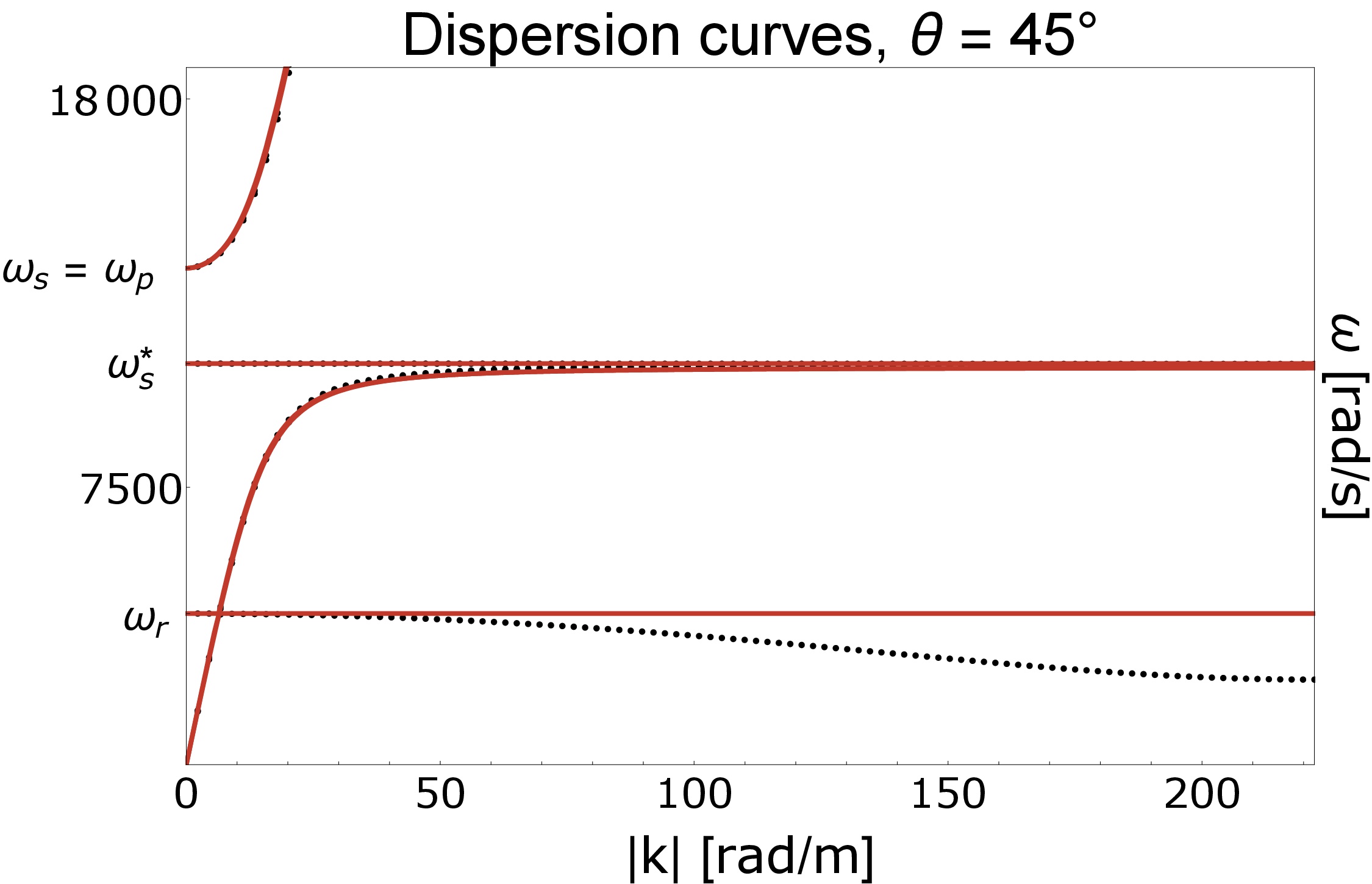}
		\caption{}
		\label{fig:disp_0_45}
	\end{subfigure}
	\caption{Fitting of the dispersion curves for the metamaterial \textit{MM1} obtained via the relaxed micromorphic model on those obtained via Bloch-Floquet analysis. The two figures correspond to different directions of propagation (a) $\theta = 0^{\circ}$, (b) $\theta = 45^{\circ}$.}
	\label{fig:disp_0_tot}
\end{figure}

This material is generated by the periodic repetition in space of the unit cell whose elastic and geometric properties are shown in Fig.~\ref{fig:fig_tab_unit_cell_0}.
We also briefly recall the values of the relaxed micromorphic parameters obtained in \cite{rizzi2020exploring} (see Table~\ref{tab:parameters_RM_0}) as the result of the dispersion curves fitting shown in Fig.~\ref{fig:disp_0_tot}.
\begin{table}[H]
	\renewcommand{\arraystretch}{1.5}
	\centering
	\begin{subtable}[t]{.70\textwidth}
		\centering
		\begin{tabular}{|c|c|c|c|c|c|c|} 
			\hline
			$\lambda_{e}$ [Pa] & $\mu_{e}$ [Pa] & $\mu^{*}_{e}$ [Pa]  & $\mu_{c}$ [Pa]  \\
			\hline
			$1.00884\times10^8$ & $2.52771\times10^9$  & $1.25592\times10^6$ & $10^5$ \\
			\hline
			$\lambda_{\rm micro}$ [Pa] & $\mu_{\rm micro}$ [Pa] & $\mu^{*}_{\rm micro}$ [Pa] & $\rho$ [kg/m$^3$]\\
			\hline
			$1.832\times10^8$  & $4.50125\times10^9$  & $2.698\times10^8$ & $3841$ \\ 
			\hline
			$L_{1}$ [m] & $L_{2}$ [m] & $L_{3}$ [m] & $L^{*}_{1}$ [m]\\
			\hline
			$0.100$ & $1.24908\times10^{-3}$ & $2.02572\times10^{-2}$ & $2.44985\times10^{-2}$\\ 
			\hline	
			$\overline{L}_{1}$ [m] & $\overline{L}_{2}$ [m] & $\overline{L}_{3}$ [m] & $\overline{L}^{*}_{1}$ [m]\\
			\hline
			$4.5639\times10^{-4}$ & $2.28195\times10^{-3}$ & $1.44323\times10^{-3}$ & $4.84074\times10^{-3}$\\ 
			\hline
		\end{tabular}
		\subcaption{}
	\end{subtable}
	\hfill
	\centering
	\begin{subtable}[t]{.20\textwidth}
		\centering
		\begin{tabular}{|c|c|c|} 
			\hline
			$\lambda_{\rm macro}$ [Pa]\\
			\hline
			$6.507\times10^7$\\
			\hline
			$\mu_{\rm macro}$ [Pa]\\
			\hline
			$1.619\times10^9$\\
			\hline
			$\mu^{*}_{\rm macro}$ [Pa]\\
			\hline
			$1.250\times10^6$\\
			\hline
		\end{tabular}
		\vspace{1.1cm}
		\subcaption{}
	\end{subtable}
	\caption{
		Panel (a) shows the values of the relaxed micromorphic static and dynamic parameters for the metamaterial \textit{MM1} determined via the fitting procedure given in \cite{d2019effective,aivaliotis2020frequency}. The apparent density $\rho$ is computed based on the titanium microstructure of Fig.~\ref{fig:fig_tab_unit_cell_0}.
		Panel (b) shows the values of the equivalent Cauchy continuum elastic coefficients corresponding to the long-wave limit of \textit{MM1} computed with the procedure explained in \cite{neff2019identification}.
	}
	\label{tab:parameters_RM_0}
\end{table}
\section{Metastructure's refractive behaviour}
\begin{figure}[H]
	\centering
	\begin{subfigure}[H]{0.45\textwidth}
	\includegraphics[width=\textwidth]{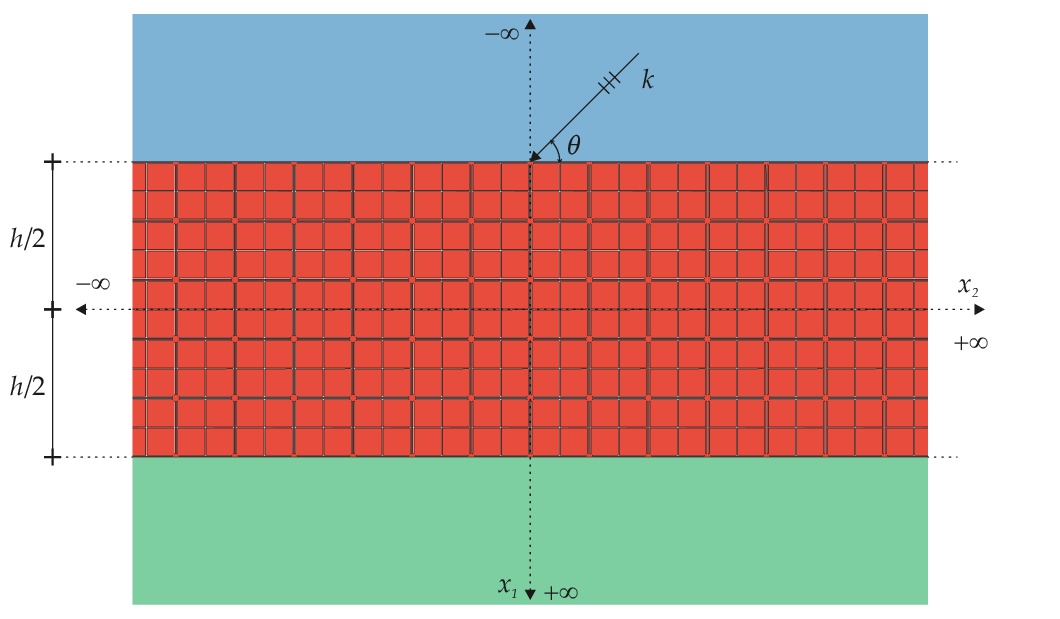}
		\caption{}
		\label{fig:slab_stru_a}
	\end{subfigure}
    \hfill
	\begin{subfigure}[H]{0.45\textwidth}
	\includegraphics[width=\textwidth]{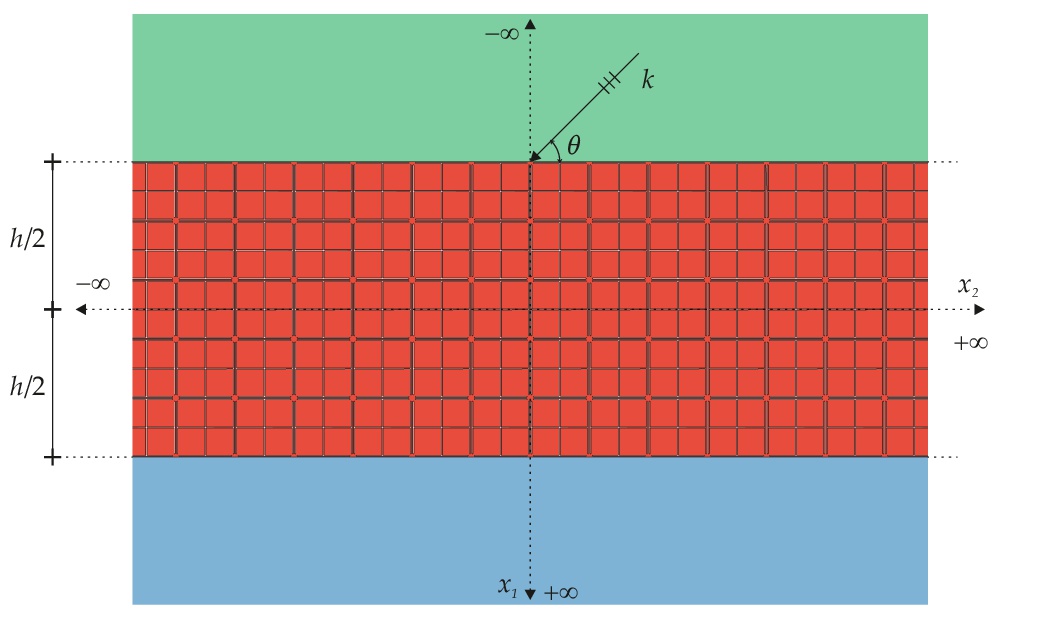}
		\caption{}
		\label{fig:slab_stru_b}
	\end{subfigure}
	\caption{Schematic representation of a wave with wavenumber $k$ hitting at angle $\theta$ a microstructured material slab of thickness $h$ embedded between two different isotropic Cauchy media.}
	\label{fig:slab_stru}
\end{figure}

In this section we will show how the relaxed micromorphic model can be suitably used to describe the refractive properties of a metamaterial's slab embedded between two different homogeneous materials (see Fig.~\ref{fig:slab_stru}).

We also show that the fact of reversing the direction of propagation of the incident wave can significantly change the structure's response in the whole range of considered frequencies.

\begin{figure}[H]
\centering
\begin{subfigure}[H]{0.37\textwidth}
	\includegraphics[width=\textwidth]{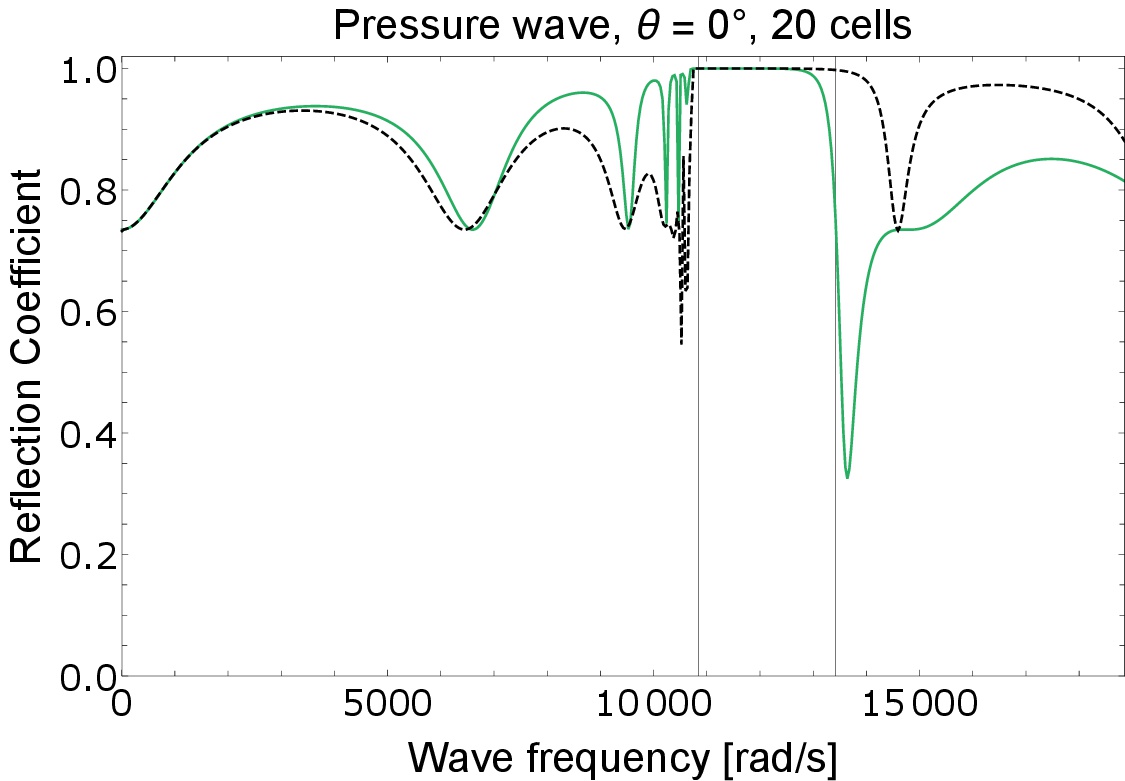}
	\caption{}
\end{subfigure}
\hspace{1cm}
\begin{subfigure}[H]{0.37\textwidth}
	\includegraphics[width=\textwidth]{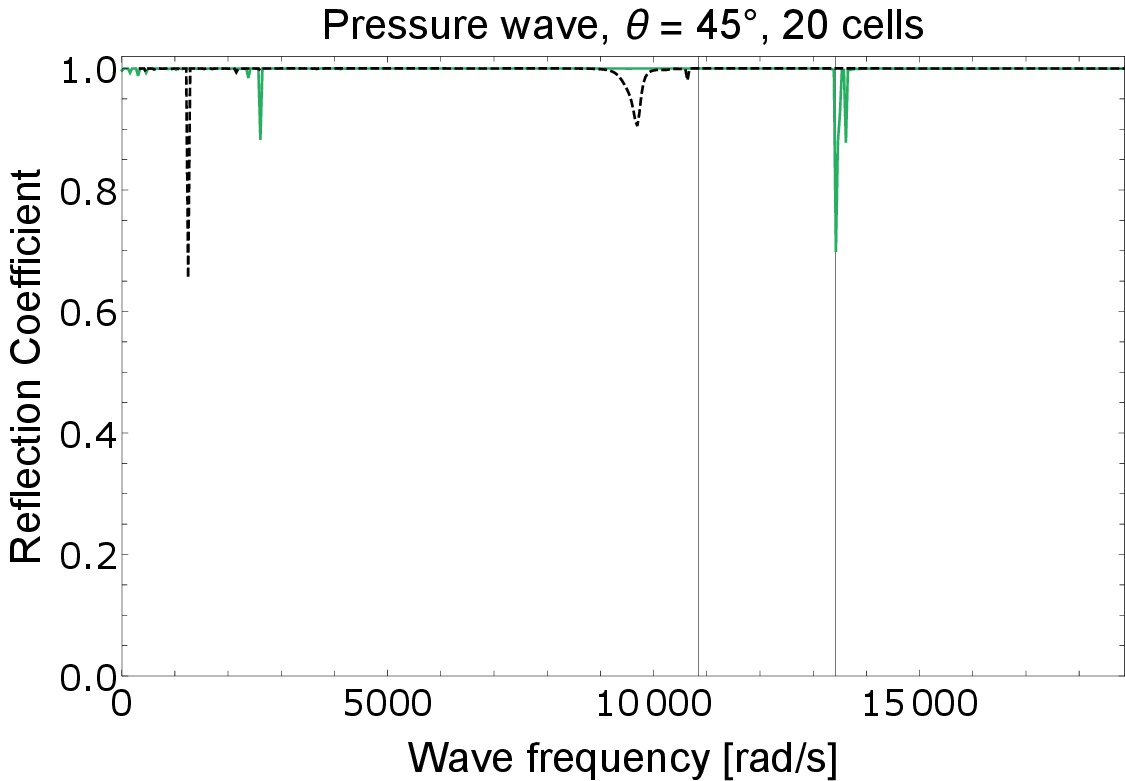}	
	\caption{}
\end{subfigure}
\hspace{1cm}
\begin{subfigure}[H]{0.37\textwidth}
	\includegraphics[width=\textwidth]{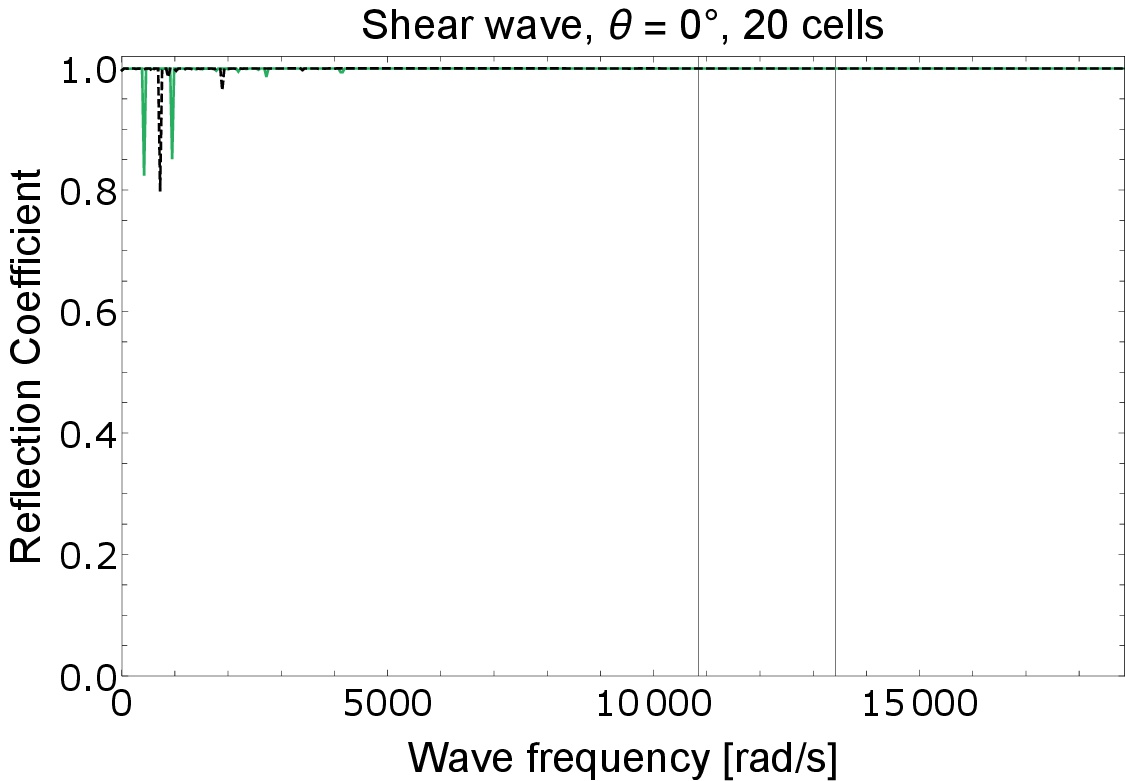}
	\caption{}
\end{subfigure}
\hspace{1cm}
\begin{subfigure}[H]{0.37\textwidth}
	\includegraphics[width=\textwidth]{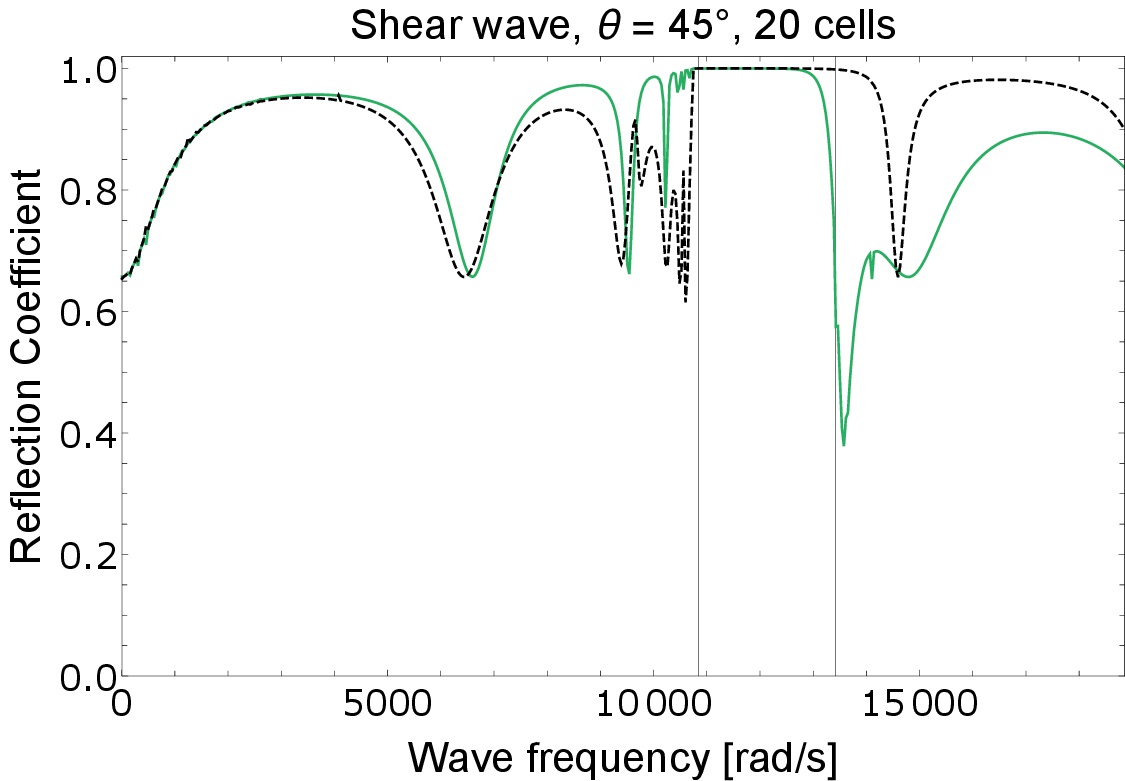}
	\caption{}
\end{subfigure}
\caption{Comparison of the microstructure's (black line) and micromorphic (green line) reflection coefficient as a function of frequency for a 20 unit cells slab of \textit{MM1} embedded between the \textit{CM1} Cauchy (blue in Fig.~\ref{fig:slab_stru_a}) and \textit{CM2} (green in Fig.~\ref{fig:slab_stru_a}).
(a) ``Pressure'' normal incident wave with respect to the slab's interface.
(b) ``Pressure'' 45$^{\circ}$ incident wave with respect to the slab's interface.
(c) ``Shear'' incident wave normal to the slab's interface.
(d) ``Shear'' 45$^{\circ}$ incident wave with respect to the slab's interface.}
\label{fig:20_cell}
\end{figure}

\begin{figure}[H]
\centering
\begin{subfigure}[H]{0.37\textwidth}
	\includegraphics[width=\textwidth]{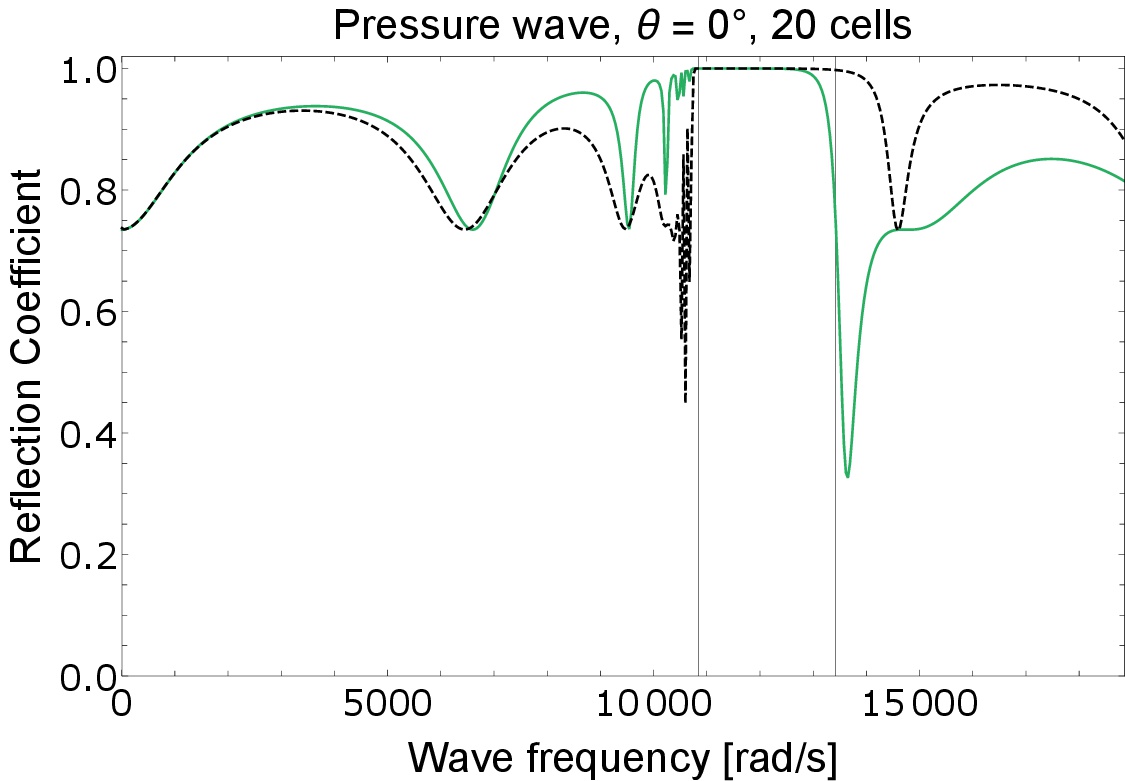}
	\caption{}
\end{subfigure}
\hspace{1cm}
\begin{subfigure}[H]{0.37\textwidth}
	\includegraphics[width=\textwidth]{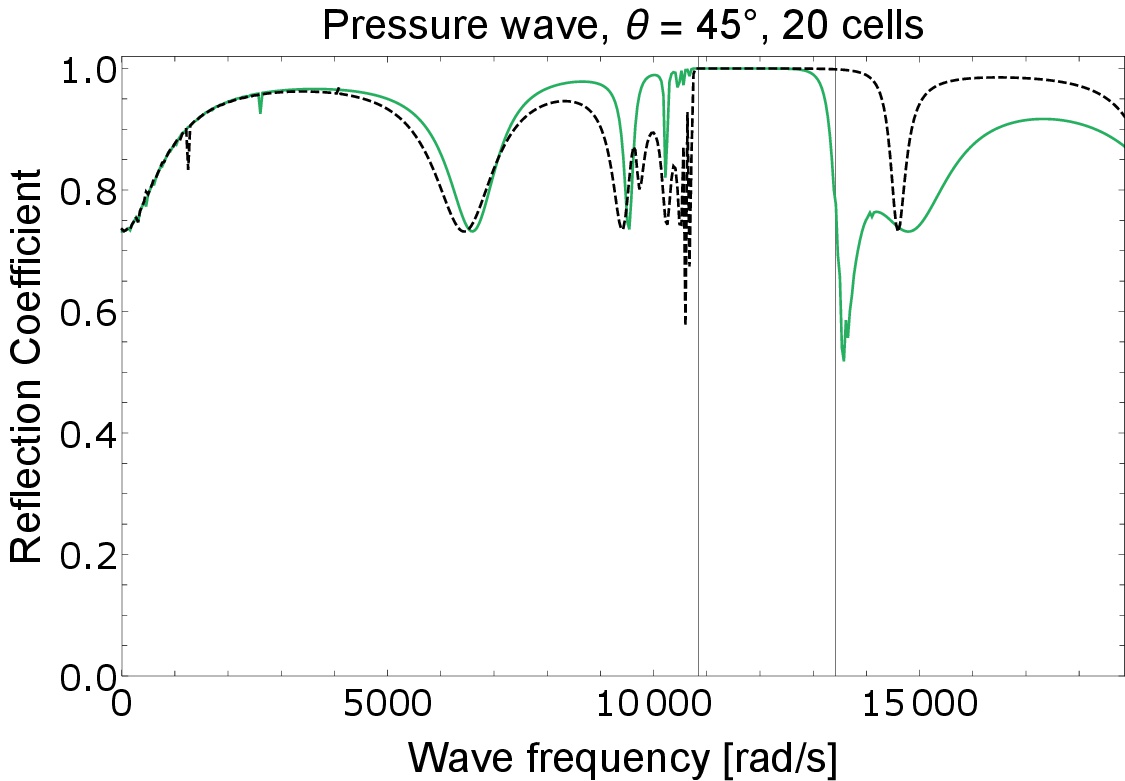}	
	\caption{}
\end{subfigure}
\hspace{1cm}
\begin{subfigure}[H]{0.37\textwidth}
	\includegraphics[width=\textwidth]{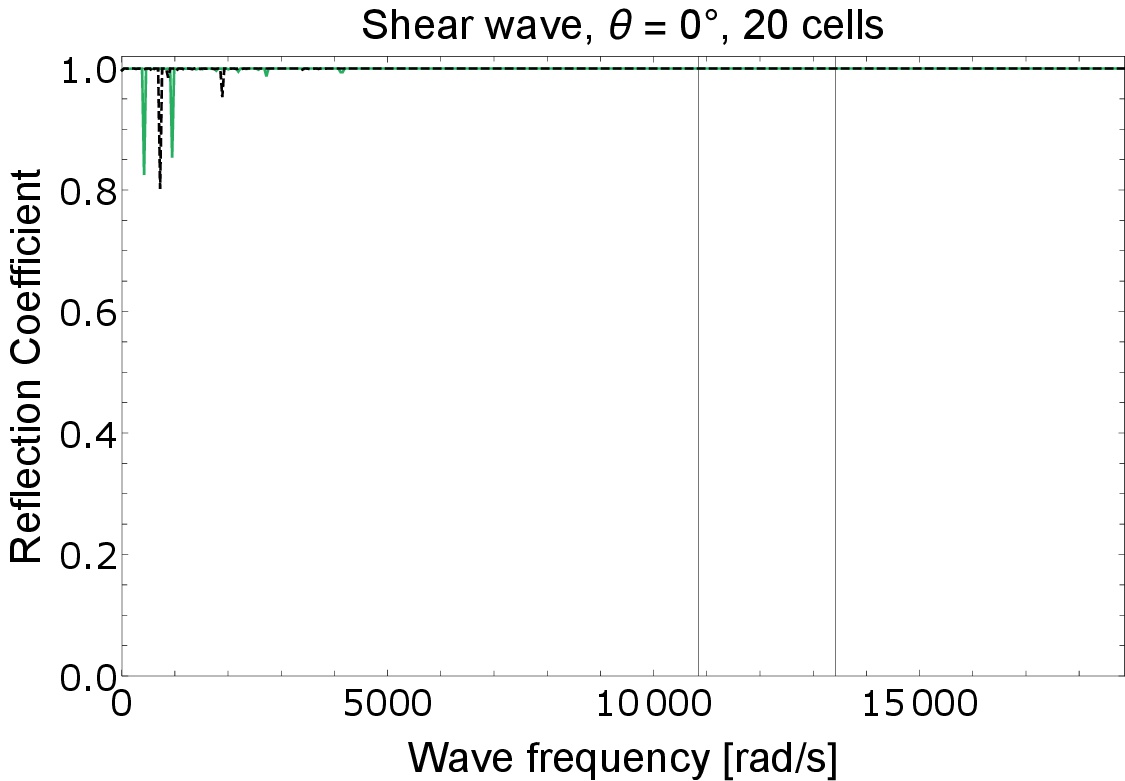}
	\caption{}
\end{subfigure}
\hspace{1cm}
\begin{subfigure}[H]{0.37\textwidth}
	\includegraphics[width=\textwidth]{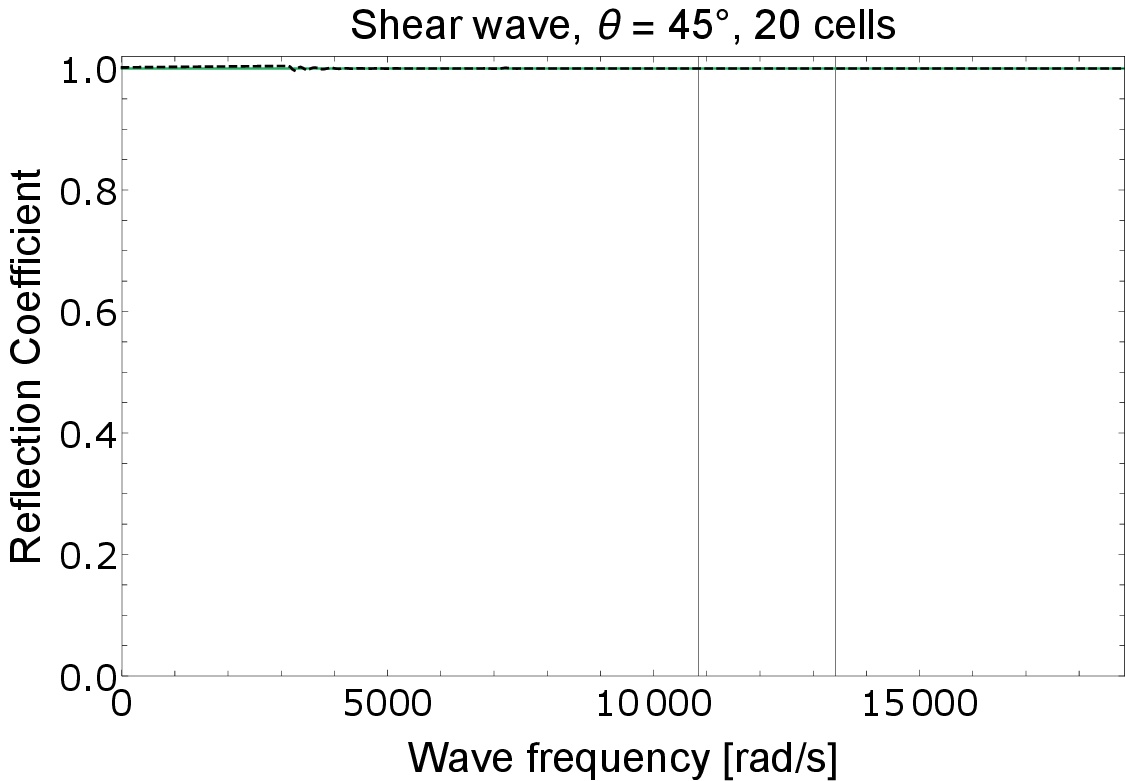}
	\caption{}
\end{subfigure}
\caption{Comparison of the microstructure's (black line) and micromorphic (green line) reflection coefficient as a function of frequency for a 20 unit cells slab of \textit{MM1} embedded between the \textit{CM2} Cauchy (green in Fig.~\ref{fig:slab_stru_b}) and \textit{CM1} (blue in Fig.~\ref{fig:slab_stru_b}).
	(a) ``pressure'' normal incident wave with respect to the slab's interface.
	(b) ``pressure'' 45$^{\circ}$ incident wave with respect to the slab's interface.
	(c) ``shear'' incident wave normal to the slab's interface.
	(d) ``shear'' 45$^{\circ}$ incident wave with respect to the slab's interface.}
\label{fig:20_cell_P_-1}
\end{figure}

Our meta-structure's design starts by the simple observation that if the two external Cauchy materials have a suitable difference in stiffness then the refractive behaviour will be different whether the incident wave travels in the top or in the bottom Cauchy material.
Due to the computational performances of the relaxed micromorphic simulations, we are able to quickly identify two Cauchy materials \textit{CM1} and \textit{CM2} for which the structure in Fig.~\ref{fig:slab_stru} acts as a diode for shear waves (Fig.~\ref{fig:20_cell} and Fig.~\ref{fig:20_cell_P_-1}).

We start by showing the reflection coefficient for the meta-structure of Fig.4 in the two cases for which the incident wave travels on one Cauchy material or in the other.
To this aim, we report the elastic properties of the two used Cauchy materials \textit{CM1} and \textit{CM2} (see Table~\ref{tab:parameters_Cau_ext_old}).

The values of the Lamé parameters are always chosen in such a way that they respect the plane strain conditions (\ref{eq:pos_def_rmm}) for positive definiteness of the strain energy.\footnote{We underline that, as it is well known, the conditions for definite positiveness under plain-strain hypothesis are $\mu>0$ and $\lambda+\mu>0$ in terms of plain strain parameters, instead of the full 3D conditions $\mu>0$ and $\lambda+\frac{2}{3}\mu>0$.}

\begin{table}[H]
	\renewcommand{\arraystretch}{1.5}
	\centering
		\begin{tabular}{|c|c|c|c|c|} 
			\hline
			$\mu_{\rm CM1}$ [Pa] & $\lambda_{\rm CM1}$ [Pa] & $\kappa_{\rm CM1}$ [Pa] & $\nu_{\rm CM1}$ [-] & $\rho_{\rm CM1}$ [kg/m$^3$]\\
			\hline
			$1.32\times10^{10}$ & $-1.31\times10^{10}$ & $10^8$ & $-0.98$ & $4400$\\
			\hline
			\hline
			$\mu_{\rm CM2}$ [Pa] & $\lambda_{\rm CM2}$ [Pa] & $\kappa_{\rm CM2}$ [Pa] & $\nu_{\rm CM2}$ [-] & $\rho_{\rm CM2}$ [kg/m$^3$]\\
			\hline
			$0.32\times10^{10}$ & $0.68\times10^{10}$ & $10^{10}$ & 0.52 & $4400$\\
			\hline
		\end{tabular}
	\caption{Values of the density, the Lamé constants, Poisson's ratio and bulk modulus under the plane strain hypothesis for (\textit{top}) the isotropic Cauchy material \textit{CM1} and (\textit{bottom}) the isotropic Cauchy material \textit{CM2}.}
	\label{tab:parameters_Cau_ext_old}
\end{table}
The reflection coefficient of the considered meta-structure is derived using eq.(\ref{eq:ref_trans_coeff})$_1$ for both the micromorphic simulation and for the microstructured finite element simulation which take into account all the geometrical details.

The PDEs (\ref{eq:equiCau}) and (\ref{eq:equiMic}) together with the interface conditions (\ref{eq:jumpdisplslab}) and (\ref{eq:jumptractionslab}) were solved semi-analytically with the software Mathematica by introducing a time harmonic plane-wave ansatz for the unknown fields $u$ and $P$.
The found solution was then used to compute the slabs's reflection coefficient by means of eq.(\ref{eq:ref_trans_coeff}).\footnote{
Since we compute the reflection coefficient on the side of the Cauchy material in which the incident wave also propagates, eq.(\ref{eq:ref_trans_coeff})$_1$ is in principle sufficient to obtain a measure of the energy which is reflected by the metamaterial's slab.
However, for the sake of completeness and to provide an independent check of our calculations, we also evaluate the transmission coefficient in the other Cauchy continuum and check, a posteriori, that the condition $\mathcal{R}+\mathcal{T}=1$ is always verified.
}

The reduced relaxed micromorphic model's structure allowed us to explore different materials and configurations so as to achieve the meta-structure that we present in the present paper.
After having selected a particular meta-structure by using the relaxed micromorphic model, we check ``a posteriori'' that its behaviour correctly corresponds to the real structure by implementing finite element simulations including all microstructural details.
Due to the periodicity of the problem in the $x_2$-direction, only a stripe of one cell thickness in the $x_1$-direction is modelled in the finite element simulation, and periodic boundary conditions are applied on the resulting left and right side.
The internal boundaries of the cross shaped holes are traction free, and to take into account the unboundness in the $x_1$-direction of the two isotropic Cauchy materials, they are modelled as two finite rectangle with a “perfectly matched layer” (PML) at the extremity.
This artificial layer dissipates the scattered field that goes through it, mimicking the response of a semi-infinite domain.
The metamaterial's slab is modelled via classical Cauchy elasticity, the base material being the one whse characteristics are reported in Fig.~\ref{fig:fig_tab_unit_cell_0}(b).

\begin{figure}[H]
\centering
\begin{subfigure}[H]{0.45\textwidth}
	\includegraphics[width=\textwidth]{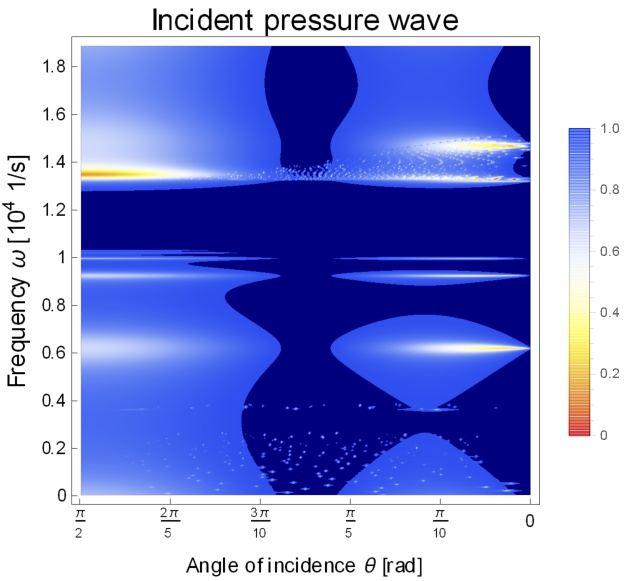}	
	\caption{}
\end{subfigure}
\hspace{1cm}
\begin{subfigure}[H]{0.45\textwidth}
	\includegraphics[width=\textwidth]{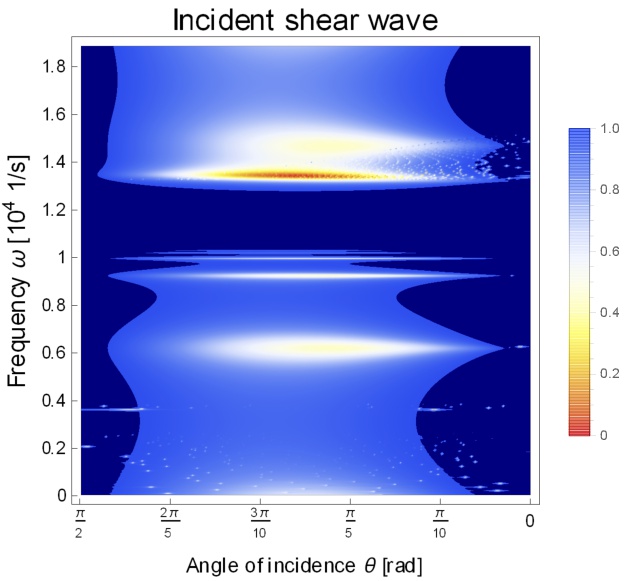}
	\caption{}
\end{subfigure}
\caption{Analytical plot of the micromorphic reflection coefficient for a 20 unit cells thick slab made up of \textit{MM1} material and embedded between the \textit{CM1} Cauchy (blue in Fig.~\ref{fig:slab_stru_a}) and \textit{CM2} (green in Fig.~\ref{fig:slab_stru_a}) as function of the angle of incidence and of the wave-frequency - (left): incident pressure wave; (right) incident shear wave.}
\label{fig:sweep_20_cell_neg_a}
\end{figure}

\begin{figure}[H]
\centering
\begin{subfigure}[H]{0.45\textwidth}
	\includegraphics[width=\textwidth]{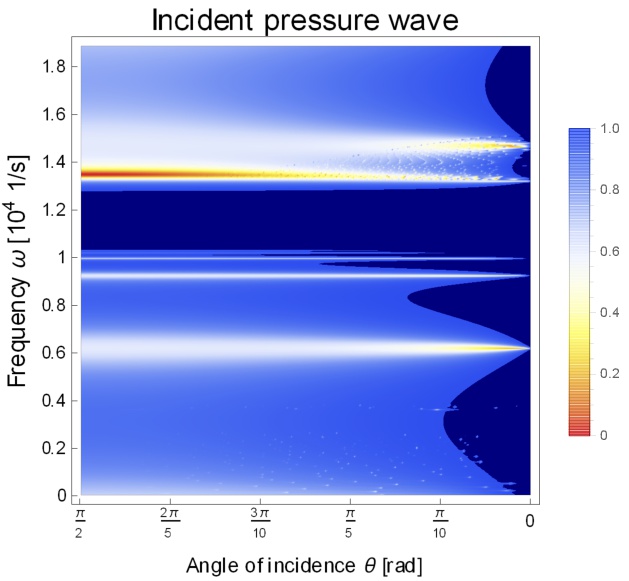}
	\caption{}	
\end{subfigure}
\hspace{1cm}
\begin{subfigure}[H]{0.45\textwidth}
	\includegraphics[width=\textwidth]{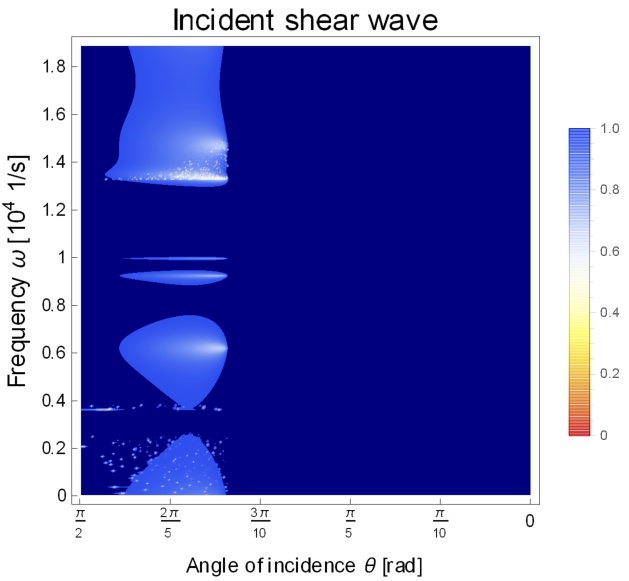}
	\caption{}
\end{subfigure}
\caption{Analytical plot of the micromorphic reflection coefficient for a 20 unit cells thick slab made up of \textit{MM1} material and embedded between the \textit{CM2} Cauchy (green in Fig.~\ref{fig:slab_stru_b}) and \textit{CM1} (blue in Fig.~\ref{fig:slab_stru_b}) as function of the angle of incidence and of the wave-frequency - (left): incident pressure wave; (right) incident shear wave.}
\label{fig:sweep_20_cell_neg_b}
\end{figure}

In this microstructured simulation the interface condition between the Cauchy media and the metamaterial are the classical ones (continuity of displacement and of Cauchy tractions).
Figures \ref{fig:20_cell} and \ref{fig:20_cell_P_-1} show that the relaxed micromorphic model describes well the refractive behaviour of the considered structure for frequencies up to the upper band-gap limit and for all the considered directions of propagation.
Thanks to the reduced relaxed micromorphic model's structure we were able to test different variants of the meta-structure shown in Fig.~\ref{fig:slab_mic} (changing the relative stiffness between the two Cauchy materials) and find the configuration shown in the present section.
As it results clearly form Fig.~\ref{fig:20_cell}, \ref{fig:20_cell_P_-1}, \ref{fig:sweep_20_cell_neg_a}, and \ref{fig:sweep_20_cell_neg_b}, this meta-structure's configuration allows low-medium frequency transmission when both pressure and shear incident waves come from the \textit{CM1} side, while transmission is almost completely prevented when a shear incident wave comes from the \textit{CM2} side.

Analyzing \textit{a priori} the complex interactions occurring between the two Cauchy materials and the metamaterial's slab so as to engineer this diode is a complex task.
However, it can be understood that, acting on the relative stiffness between the two external Cauchy materials, may affect the meta-structure's behaviour to a significant extent.
Based on this simple remark, we exploited the relaxed micromorphic model computational performances in order to test many combinations of relative stiffness between \textit{CM1} and \textit{CM2}, in order to maximize the reflected energy for a large range of angles of incidence when considering a shear wave.
We quickly ended up with the structure whose performances are shown in Fig \ref{fig:20_cell}-\ref{fig:sweep_20_cell_neg_b}.
This meta-structure thus acts as a diode for shear-incident waves for a very wide range of frequencies and angles of incidence.
In the reminder of this paper, we will show that this behaviour can be enhanced and optimized by simply acting on the relative stiffness of the two Cauchy materials, thus giving rise to diodes both for pressure and shear incident waves.
\section{Optimization of the considered meta-structure: conception of an effective low-medium-frequency diode/high-frequency screen}
Thanks to the computational advantages given by the relaxed micromorphic model’s reduced structure, we can efficiently explore a wide range of stiffnesses around the values given in Table~\ref{tab:parameters_Cau_ext_old} for the two Cauchy materials. This allowed us to find an optimized combination such that the structure acts as a diode for low-medium frequencies for both pressure and shear incident waves and for a wide range of angles of incidence.
\begin{table}[H]
	\renewcommand{\arraystretch}{1.5}
	\centering
		\begin{tabular}{|c|c|c|c|c|} 
			\hline
			$\mu_{\rm CM3}$ [Pa] & $\lambda_{\rm CM3}$ [Pa] & $\kappa_{\rm CM3}$ [Pa] & $\nu_{\rm CM3}$ [-] & $\rho_{\rm CM3}$ [kg/m$^3$]\\
			\hline
			$5.52\times10^{10}$ & $-5.33\times10^{10}$ & $1.9\times10^9$ & $-0.93$ & $4400$\\
			\hline
			\hline
			$\mu_{\rm CM4}$ [Pa] & $\lambda_{\rm CM4}$ [Pa] & $\kappa_{\rm CM4}$ [Pa] & $\nu_{\rm CM4}$ [-] & $\rho_{\rm CM4}$ [kg/m$^3$]\\
			\hline
			$2.66\times10^{8}$ & $-2.57\times10^{8}$ & $9\times10^{6}$ & $-0.93$ & $4400$\\
			\hline
		\end{tabular}
	\caption{Values of the density, the Lamé constants, Poisson's ratio and bulk modulus under the plane strain hypothesis for (\textit{top}) the isotropic Cauchy material \textit{CM3} and (\textit{bottom}) the isotropic Cauchy material \textit{CM4}.}
	\label{tab:parameters_Cau_ext_stiffer}
\end{table}

\begin{figure}[H]
	\centering
	\begin{subfigure}[H]{0.37\textwidth}
		\includegraphics[width=\textwidth]{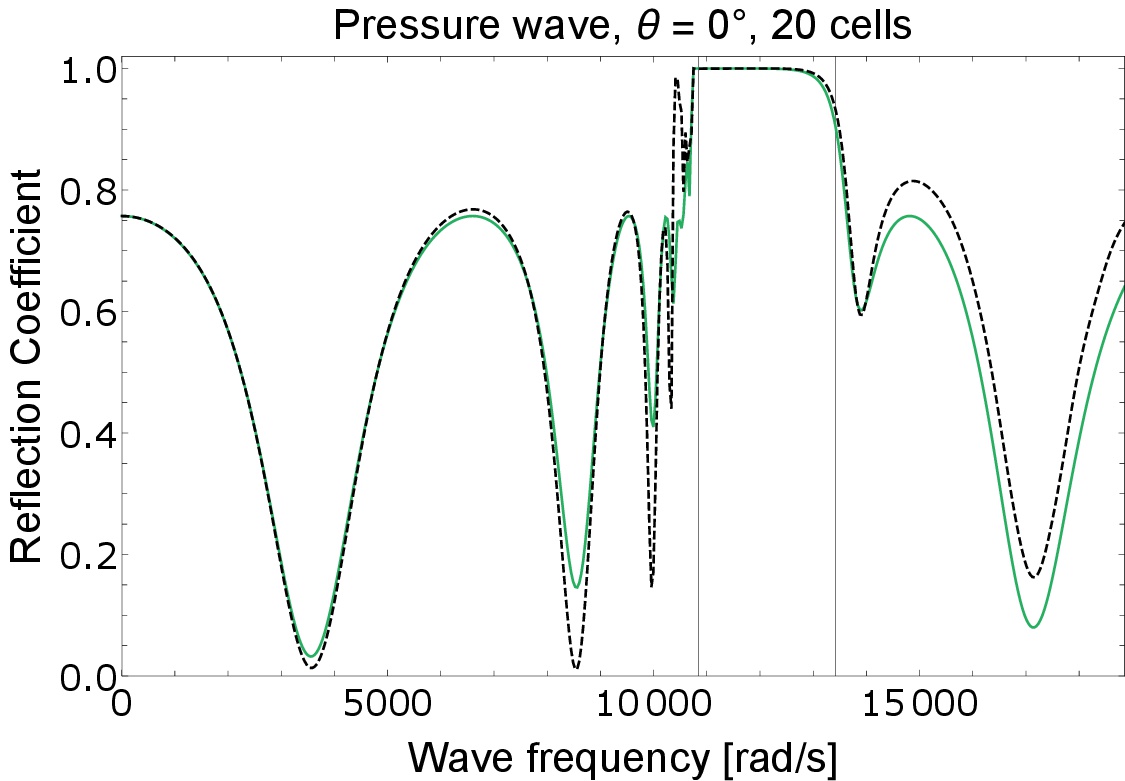}
		\caption{}
	\end{subfigure}
	\hspace{1cm}
	\begin{subfigure}[H]{0.37\textwidth}
		\includegraphics[width=\textwidth]{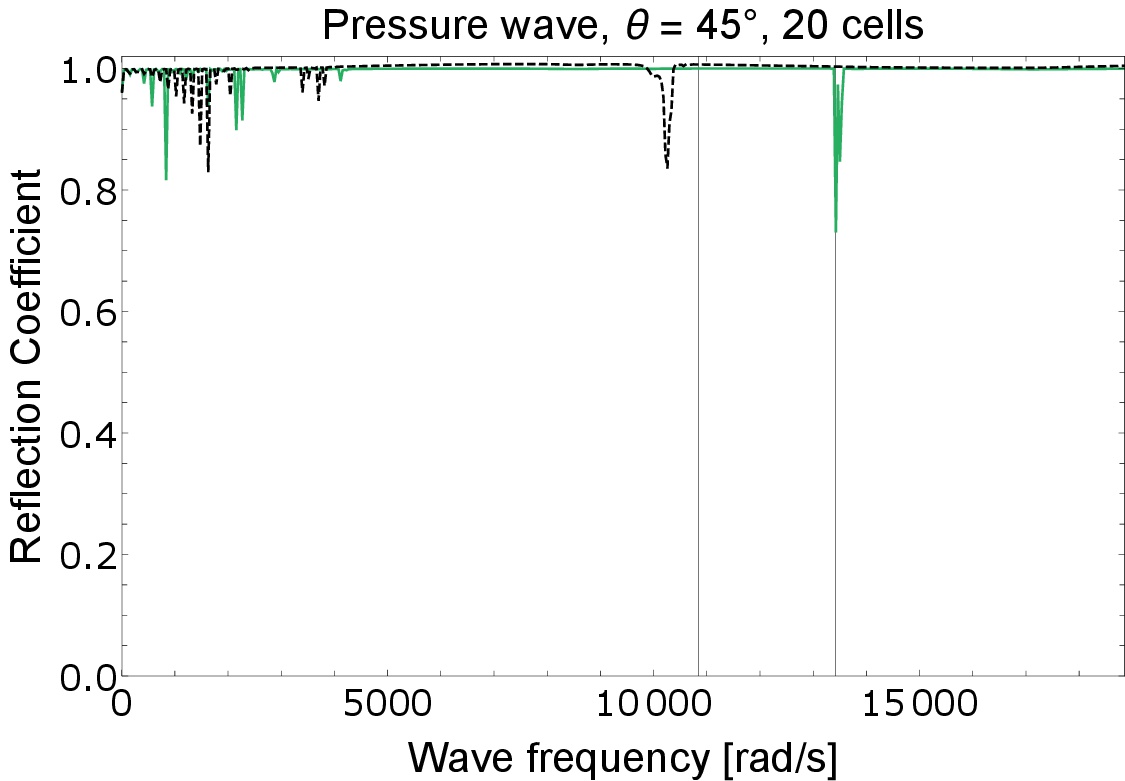}	
		\caption{}
	\end{subfigure}
	\hspace{1cm}
	\begin{subfigure}[H]{0.37\textwidth}
		\includegraphics[width=\textwidth]{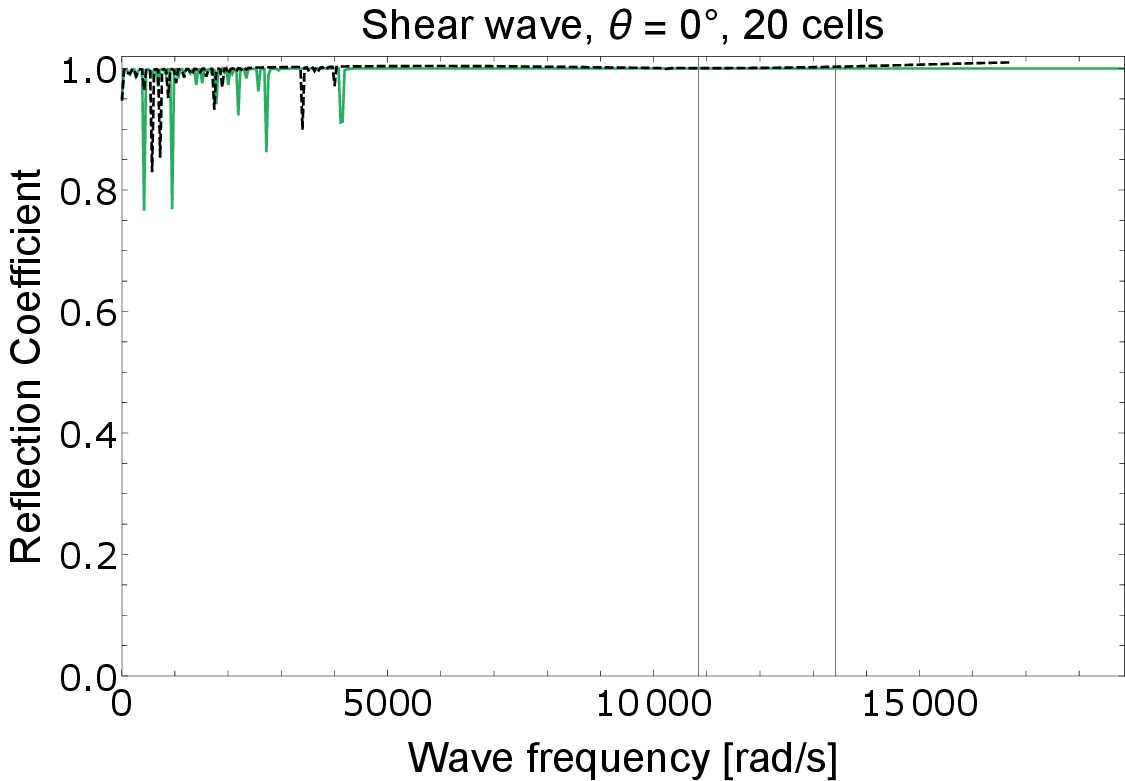}
		\caption{}
	\end{subfigure}
	\hspace{1cm}
	\begin{subfigure}[H]{0.37\textwidth}
		\includegraphics[width=\textwidth]{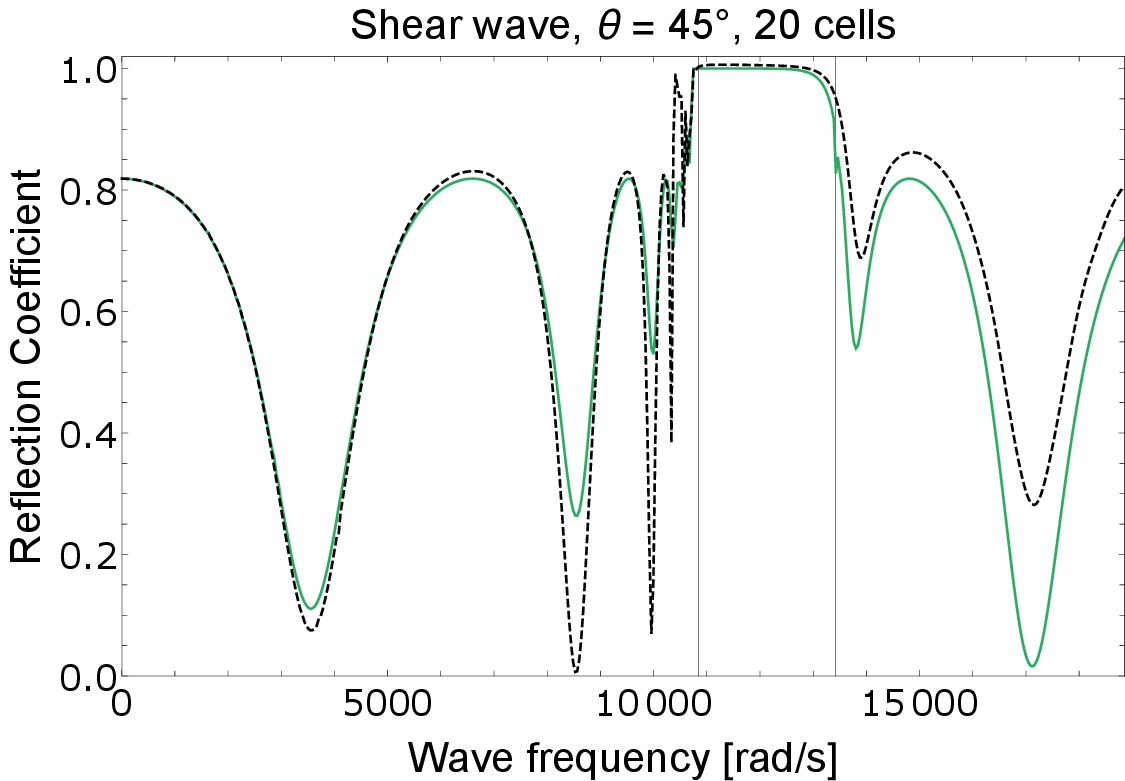}
		\caption{}
	\end{subfigure}
	\caption{Comparison of the microstructure's (black line) and micromorphic (green line) reflection coefficient as a function of frequency for a 20 unit cells slab of \textit{MM1} embedded between the \textit{CM3} Cauchy (green in Fig.~\ref{fig:slab_stru_a}) and \textit{CM4} (blue in Fig.~\ref{fig:slab_stru_a}).
		(a) ``pressure'' normal incident wave with respect to the slab's interface.
		(b) ``pressure'' 45$^{\circ}$ incident wave with respect to the slab's interface.
		(c) ``shear'' incident wave normal to the slab's interface.
		(d) ``shear'' 45$^{\circ}$ incident wave with respect to the slab's interface.}
	\label{fig:20_cell_caso_1}
\end{figure}

\begin{figure}[H]
	\centering
	\begin{subfigure}[H]{0.37\textwidth}
		\includegraphics[width=\textwidth]{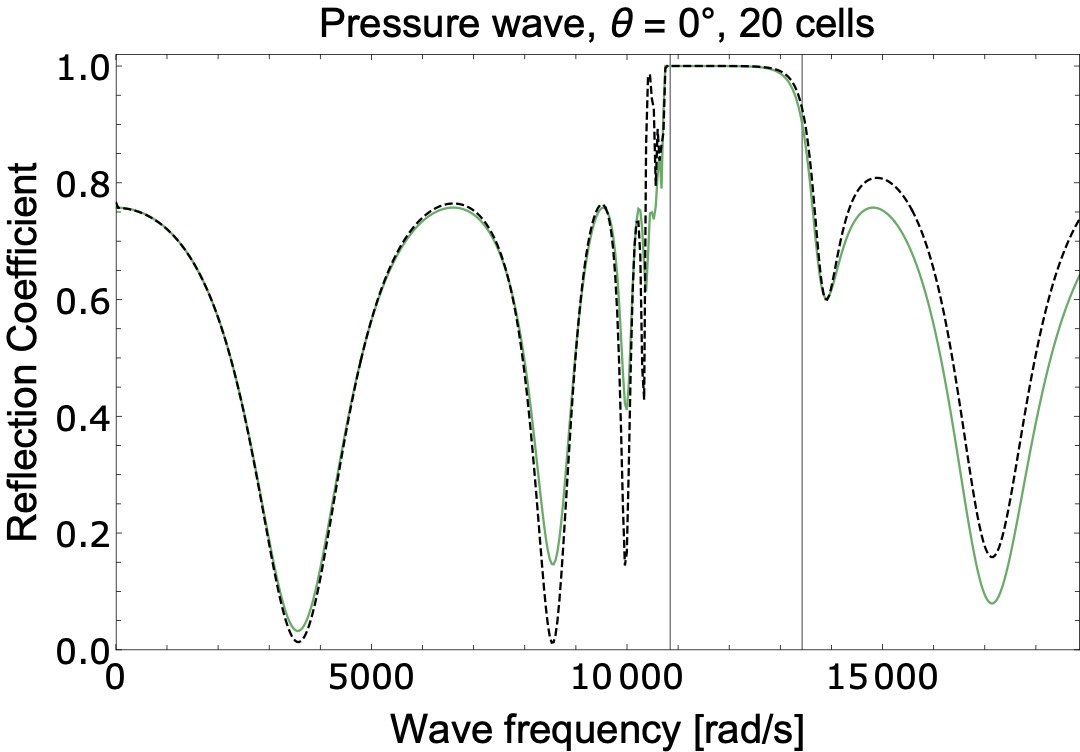}
		\caption{}
	\end{subfigure}
	\hspace{1cm}
	\begin{subfigure}[H]{0.37\textwidth}
		\includegraphics[width=\textwidth]{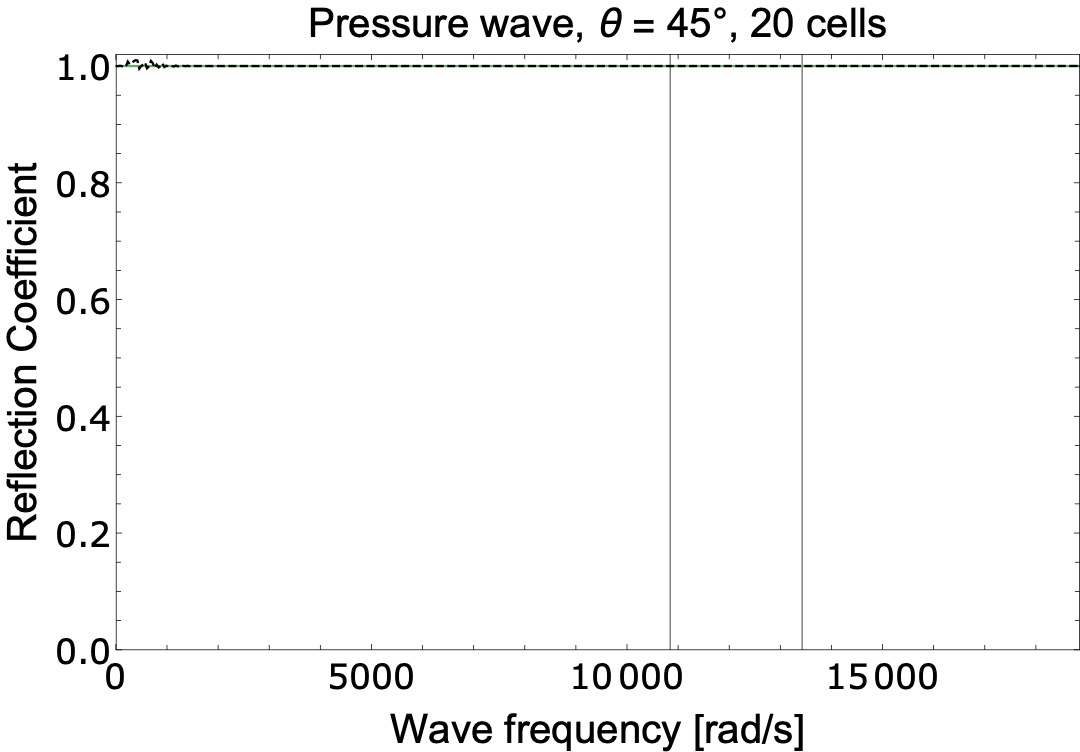}	
		\caption{}
	\end{subfigure}
	\hspace{1cm}
	\begin{subfigure}[H]{0.37\textwidth}
		\includegraphics[width=\textwidth]{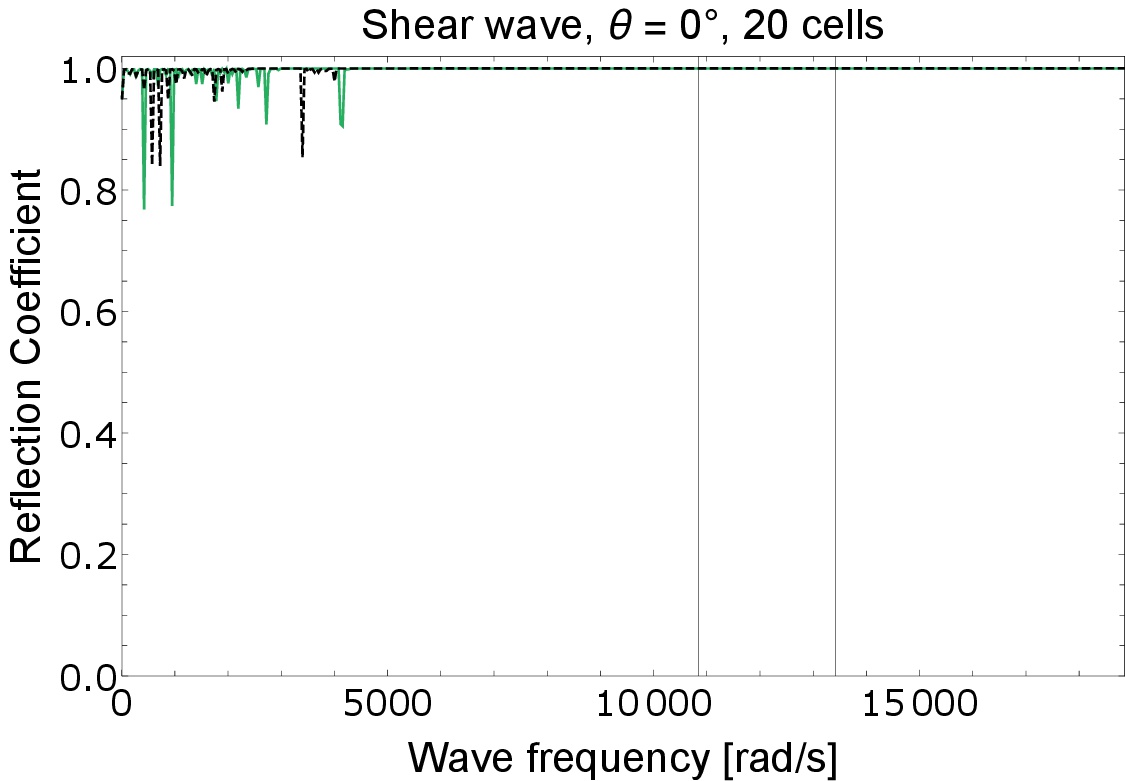}
		\caption{}
	\end{subfigure}
	\hspace{1cm}
	\begin{subfigure}[H]{0.37\textwidth}
		\includegraphics[width=\textwidth]{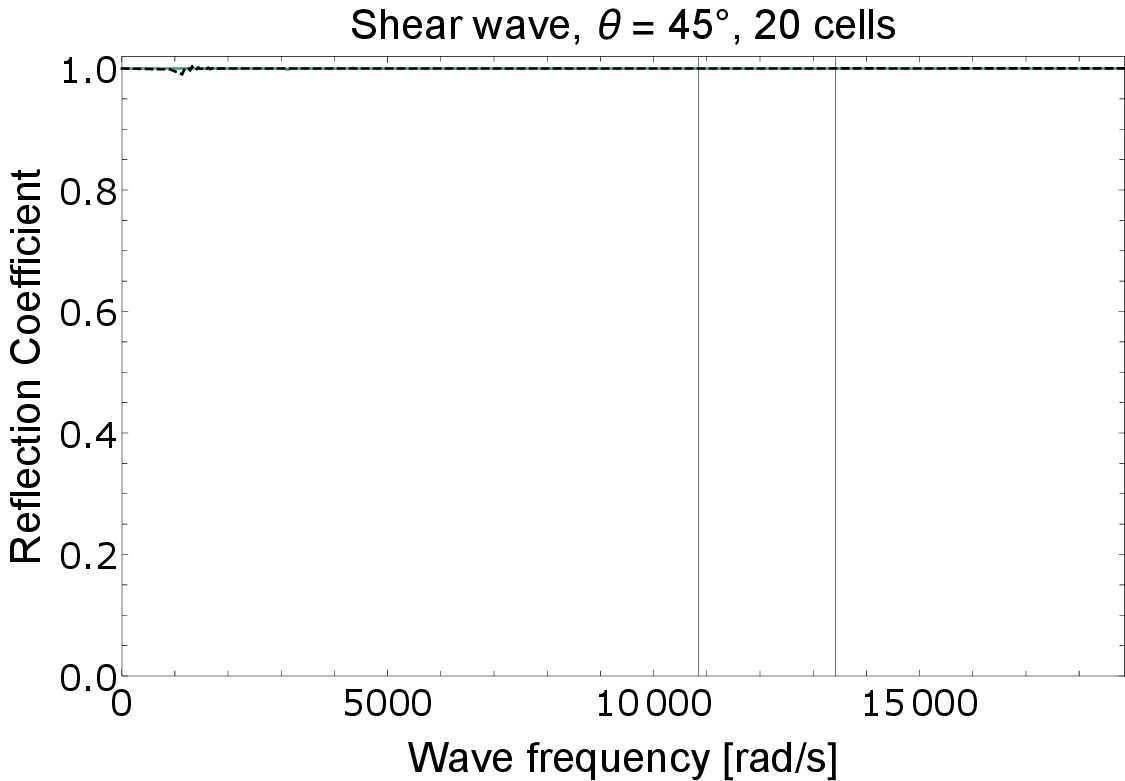}
		\caption{}
	\end{subfigure}
	\caption{Comparison of the microstructure's (black line) and micromorphic (green line) reflection coefficient as a function of frequency for a 20 unit cells slab of \textit{MM1} embedded between the \textit{CM4} Cauchy (green in Fig.~\ref{fig:slab_stru_b}) and \textit{CM3} (blue in Fig.~\ref{fig:slab_stru_b}).
		(a) ``pressure'' normal incident wave with respect to the slab's interface.
		(b) ``pressure'' 45$^{\circ}$ incident wave with respect to the slab's interface.
		(c) ``shear'' incident wave normal to the slab's interface.
		(d) ``shear'' 45$^{\circ}$ incident wave with respect to the slab's interface.}
	\label{fig:20_cell_caso_2}
\end{figure}

\begin{figure}[H]
	\centering
	\begin{subfigure}[H]{0.45\textwidth}
		\includegraphics[width=\textwidth]{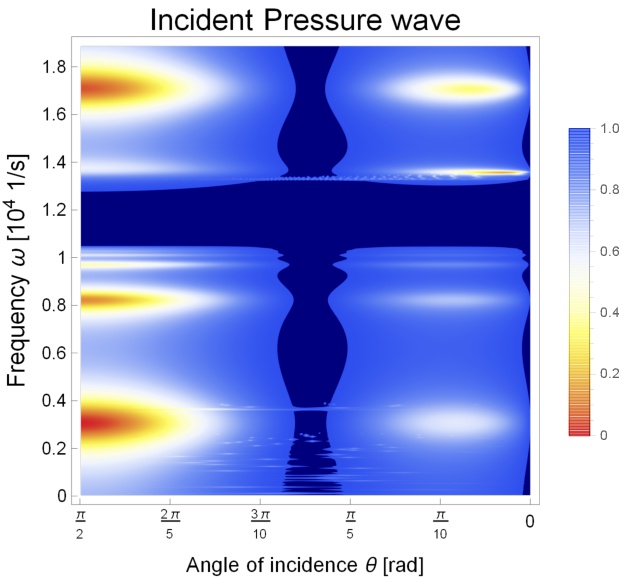}	
		\caption{}
	\end{subfigure}
	\hspace{1cm}
	\begin{subfigure}[H]{0.45\textwidth}
		\includegraphics[width=\textwidth]{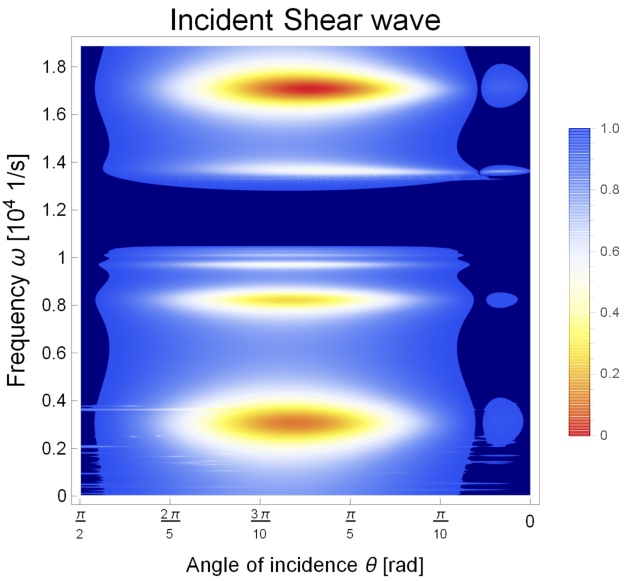}
		\caption{}
	\end{subfigure}
	\caption{Analytical plot of the micromorphic reflection coefficient for a 20 unit cells thick slab made up of \textit{MM1} material and embedded between the \textit{CM3} Cauchy (green in Fig.~\ref{fig:slab_stru_a}) and \textit{CM4} (blue in Fig.~\ref{fig:slab_stru_a}), as function of the angle of incidence and of the wave-frequency - (left): incident pressure wave; (right) incident shear wave.}
	\label{fig:sweep_20_cell_caso_1}
\end{figure}

\begin{figure}[H]
	\centering
	\begin{subfigure}[H]{0.45\textwidth}
		\includegraphics[width=\textwidth]{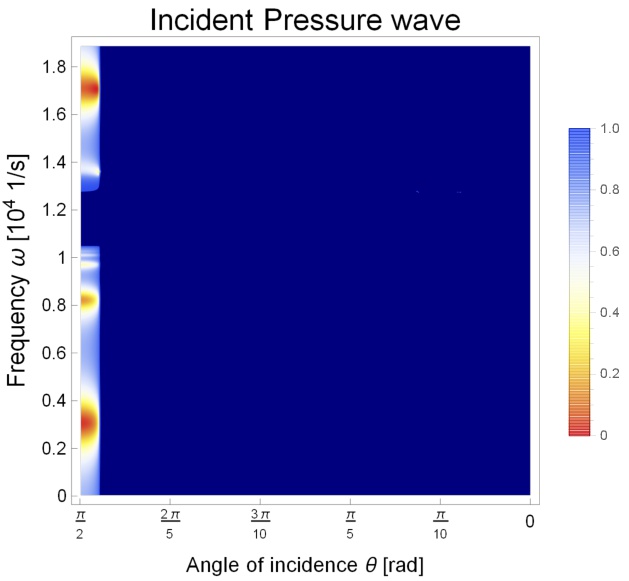}
		\caption{}	
	\end{subfigure}
	\hspace{1cm}
	\begin{subfigure}[H]{0.45\textwidth}
		\includegraphics[width=\textwidth]{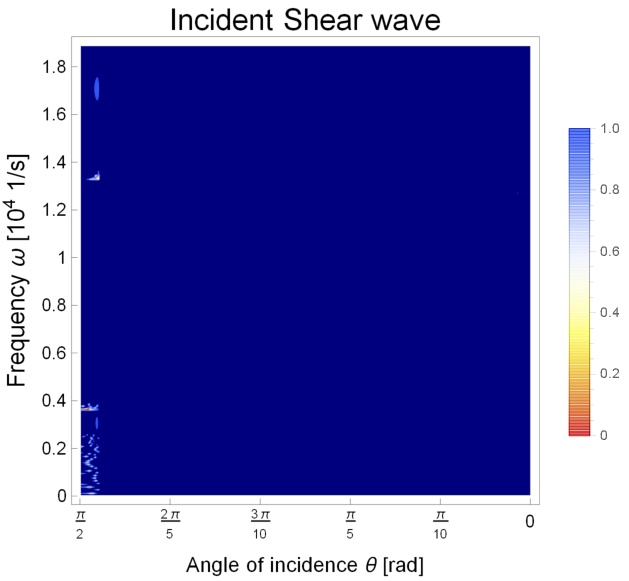}
		\caption{}
	\end{subfigure}
\caption{Analytical plot of the micromorphic reflection coefficient for a 20 unit cells thick slab made up of \textit{MM1} material and embedded between the \textit{CM4} Cauchy (green in Fig.~\ref{fig:slab_stru_b}) and \textit{CM3} (blue in Fig.~\ref{fig:slab_stru_b}), as function of the angle of incidence and of the wave-frequency - (left): incident pressure wave; (right) incident shear wave.}
	\label{fig:sweep_20_cell_caso_2}
\end{figure}

Using the elastic values given in Table~\ref{tab:parameters_Cau_ext_stiffer}, we show in Fig.~\ref{fig:20_cell_caso_1}-\ref{fig:sweep_20_cell_caso_2} that the refractive behaviour of the considered meta-structure drastically changes if the incident wave comes from the side of the ``softer'' or the ``stiffer'' material.\footnote{
The two homogeneous materials whose parameters are given in Table~\ref{tab:parameters_Cau_ext_stiffer} have been chosen so as to maximise the diode behaviour for the largest possible range of angle of incidence.
The main physical mechanism behind the diode behaviour is due to an increase (or decrease) of the stiffness between the Cauchy material on top, the metamaterial in between, and the Cauchy material on the bottom.
This means that, if a sufficient difference in stiffness is provided between the two Cauchy materials, the diode behaviour can be achieved with more "common" materials with respect those presented in Table~\ref{tab:parameters_Cau_ext_stiffer} (for example aluminum and silicon rubber).
}
In particular, figures \ref{fig:20_cell_caso_1} and \ref{fig:20_cell_caso_2}, also show that the relaxed micromorphic model catches well the meta-structure's response for a very large range of frequencies going well beyond the first band-gap.
With reference to Fig.~\ref{fig:sweep_20_cell_caso_1} and Fig.~\ref{fig:sweep_20_cell_caso_2}, we can remark that in the low-medium frequency range the structure acts as an almost perfect diode in which both pressure and shear waves are able to travel if they come from the side of the ``stiffer'' Cauchy materials, while are completely blocked if they come from the side of the ``softer'' material.

\begin{figure}[H]
	\centering
	\begin{subfigure}[H]{0.45\textwidth}
		\includegraphics[width=\textwidth]{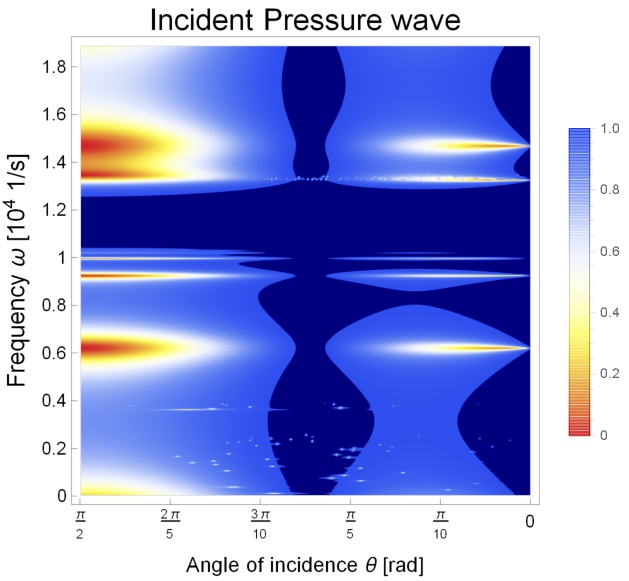}	
		\caption{}
	\end{subfigure}
	\hspace{1cm}
	\begin{subfigure}[H]{0.45\textwidth}
		\includegraphics[width=\textwidth]{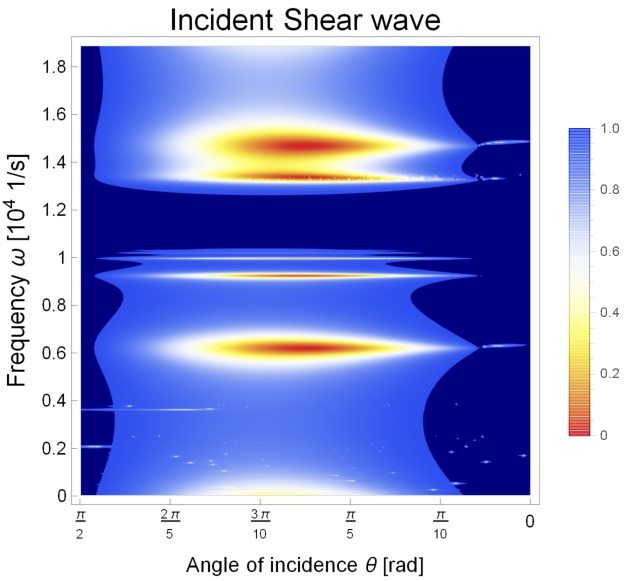}
		\caption{}
	\end{subfigure}
	\caption{Analytical plot of the micromorphic reflection coefficient for a 20 unit cells thick slab made up of \textit{MM1} material and embedded only between the \textit{CM3} Cauchy, as function of the angle of incidence and of the wave-frequency - (left): incident pressure wave; (right) incident shear wave.}
	\label{fig:sweep_20_cell_caso_a}
\end{figure}

\begin{figure}[H]
	\centering
	\begin{subfigure}[H]{0.45\textwidth}
		\includegraphics[width=\textwidth]{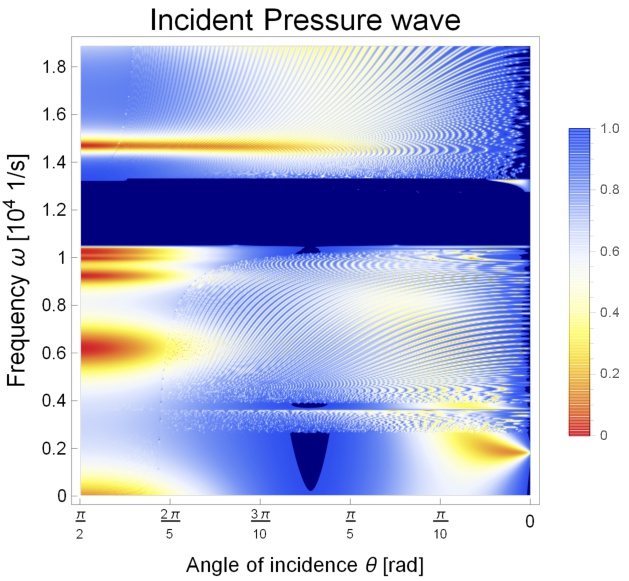}
		\caption{}	
	\end{subfigure}
	\hspace{1cm}
	\begin{subfigure}[H]{0.45\textwidth}
		\includegraphics[width=\textwidth]{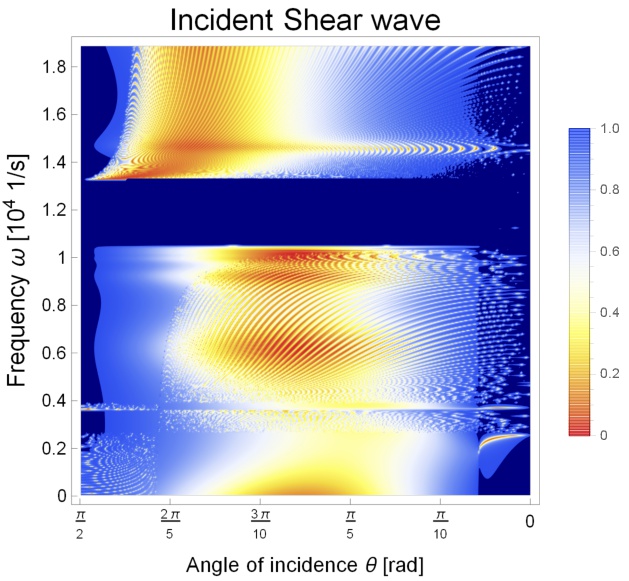}
		\caption{}
	\end{subfigure}
	\caption{Analytical plot of the micromorphic reflection coefficient for a 20 unit cells thick slab made up of \textit{MM1} material and embedded only between the \textit{CM4} Cauchy, as function of the angle of incidence and of the wave-frequency - (left): incident pressure wave; (right) incident shear wave.}
	\label{fig:sweep_20_cell_caso_b}
\end{figure}

As we already remarked, this diode behaviour is mainly due to the difference of stiffness of the two Cauchy materials given in Table~\ref{tab:parameters_Cau_ext_stiffer}.
To clearly explain this claim, we show in Fig.~\ref{fig:sweep_20_cell_caso_a} and Fig.~\ref{fig:sweep_20_cell_caso_b} the behaviour of the considered meta-structure when we embed the metamaterial's slab between two layers of the Cauchy material \textit{CM3} of Table~\ref{tab:parameters_Cau_ext_stiffer}(a), or between two layers of the Cauchy material \textit{CM4} given in Table~\ref{tab:parameters_Cau_ext_stiffer}(b), respectively.
It is clear that, when no difference of stiffness exists between the two homogeneous layers, then transmission occurs for any direction of the incident wave and for any angle of incidence as far as low-medium frequencies are considered (see Fig.~\ref{fig:sweep_20_cell_caso_a} and Fig.~\ref{fig:sweep_20_cell_caso_b})
We finally show that, even if the diode-behaviour observed at low-medium frequencies is mainly driven by the relative stiffness of the two homogenous materials, the fact of having a metamaterial's slab in the middle of them drastically changes the meta-structure's medium-high frequency response.

\begin{figure}[H]
	\centering
	\begin{subfigure}[H]{0.45\textwidth}
		\includegraphics[width=\textwidth]{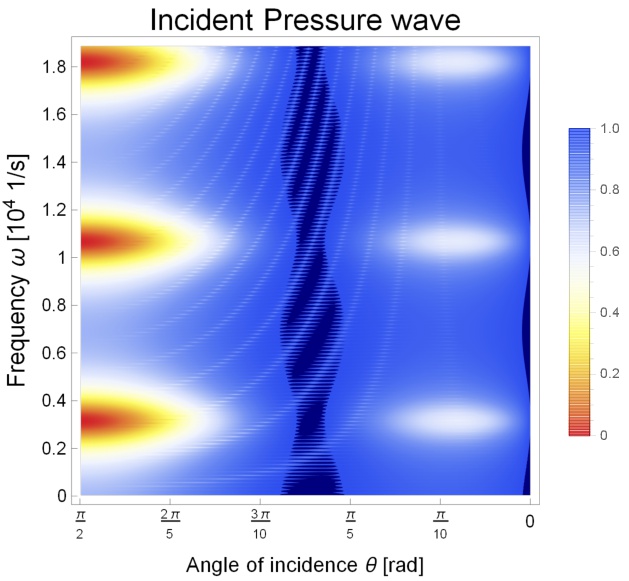}	
		\caption{}
	\end{subfigure}
	\hspace{1cm}
	\begin{subfigure}[H]{0.45\textwidth}
		\includegraphics[width=\textwidth]{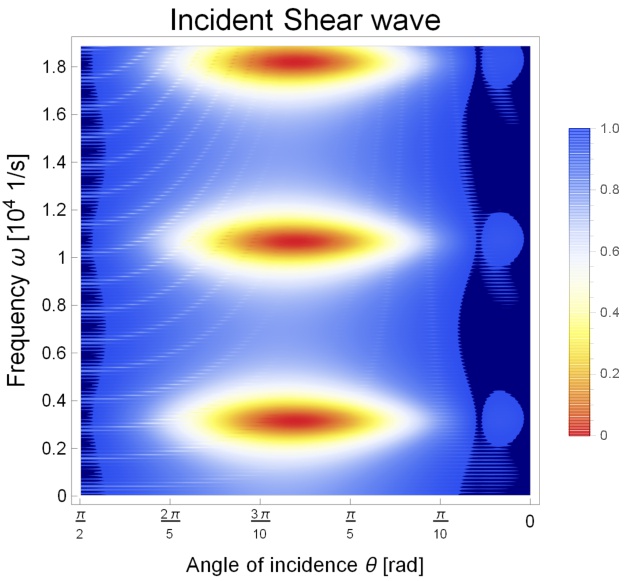}
		\caption{}
	\end{subfigure}
	\caption{Analytical plot of the effective Cauchy anisotropic material reflection coefficient for a 20 unit cells thick slab embedded between the \textit{CM3} Cauchy (green in Fig.~\ref{fig:slab_stru_a}) and \textit{CM4} (blue in Fig.~\ref{fig:slab_stru_a}), as function of the angle of incidence and of the wave-frequency - (left): incident pressure wave; (right) incident shear wave.}
	\label{fig:sweep_20_cell_caso_1_Cau}
\end{figure}

\begin{figure}[H]
	\centering
	\begin{subfigure}[H]{0.45\textwidth}
		\includegraphics[width=\textwidth]{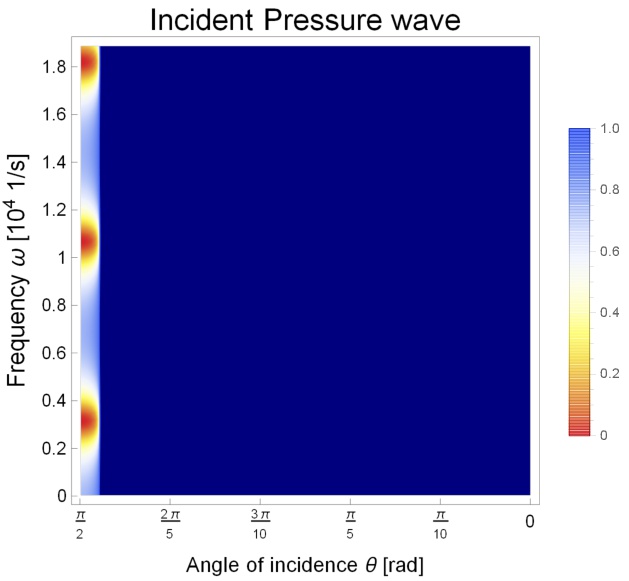}
		\caption{}	
	\end{subfigure}
	\hspace{1cm}
	\begin{subfigure}[H]{0.45\textwidth}
		\includegraphics[width=\textwidth]{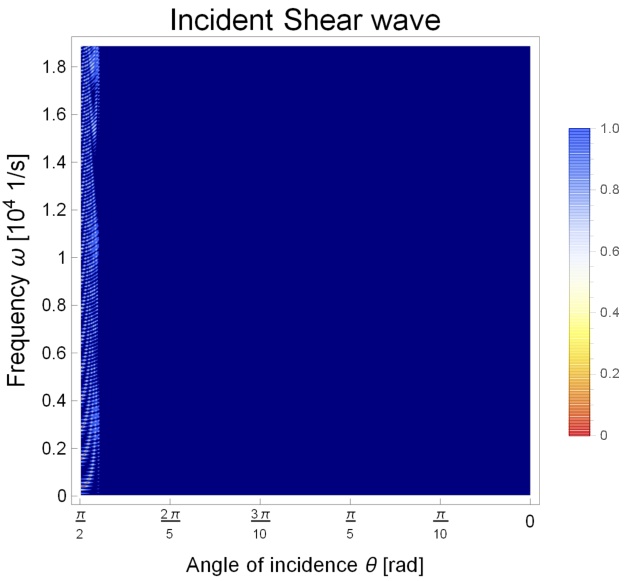}
		\caption{}
	\end{subfigure}
	\caption{Analytical plot of the effective Cauchy anisotropic material reflection coefficient for a 20 unit cells thick slab embedded between the \textit{CM4} Cauchy (green in Fig.~\ref{fig:slab_stru_b}) and \textit{CM3} (blue in Fig.~\ref{fig:slab_stru_b}), as function of the angle of incidence and of the wave-frequency - (left): incident pressure wave; (right) incident shear wave.}
	\label{fig:sweep_20_cell_caso_2_Cau}
\end{figure}

To better explain this claim, we show in Fig.~\ref{fig:sweep_20_cell_caso_1_Cau} and Fig.~\ref{fig:sweep_20_cell_caso_2_Cau} the meta-structure's refractive behaviour, when the interior slab is modelled via a tetragonal Cauchy model, instead that via the relaxed micromorphic model.

This tetragonal Cauchy material (see Table~\ref{tab:parameters_RM_0}(b)) has macroscopic stiffnesses derived as the long-wave limit of the relaxed micromorphic material presented in \ref{tab:parameters_RM_0}(a)).
It is clear from Fig.~\ref{fig:sweep_20_cell_caso_1_Cau} and Fig.~\ref{fig:sweep_20_cell_caso_2_Cau} that, if the low-frequency diode response is well described, the medium-higher frequency response is not.
This calls for the need of using the relaxed micromorphic model to well describe the meta-structure's behaviour for all the possible frequencies.

\begin{figure}[H]
	\centering
	\begin{subfigure}[H]{0.45\textwidth}
		\includegraphics[width=\textwidth]{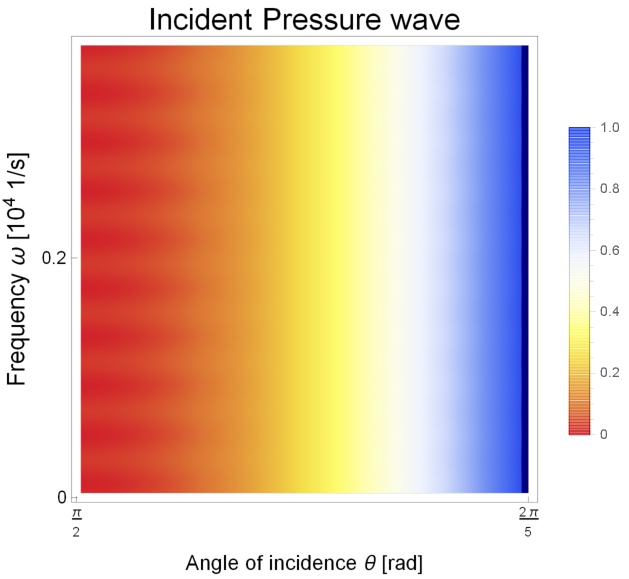}
		\caption{}	
	\end{subfigure}
	\hspace{1cm}
	\begin{subfigure}[H]{0.45\textwidth}
		\includegraphics[width=\textwidth]{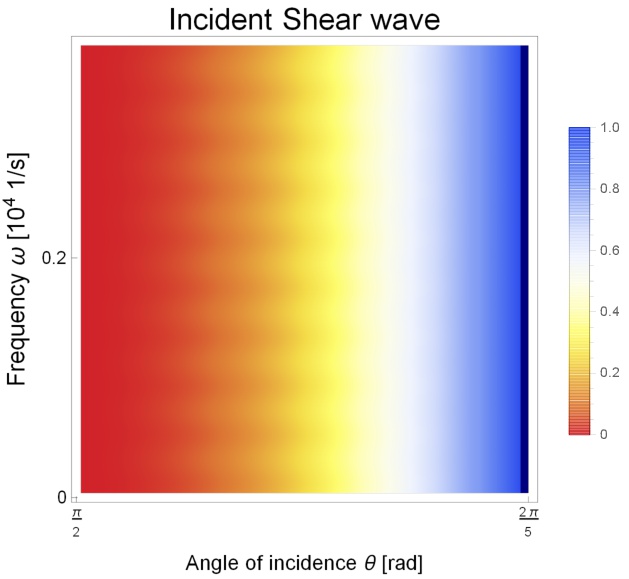}
		\caption{}
	\end{subfigure}
\caption{Analytical plot of a single interface separating two Cauchy materials \text{CM3} and \text{CM4} (no metamaterial's slab) reflection coefficient, as function of the angle of incidence and of the wave-frequency - (left): incident pressure wave; (right) incident shear wave.}
	\label{fig:sweep_20_cell_caso_a_no_slab}
\end{figure}

\begin{figure}[H]
	\centering
	\begin{subfigure}[H]{0.45\textwidth}
		\includegraphics[width=\textwidth]{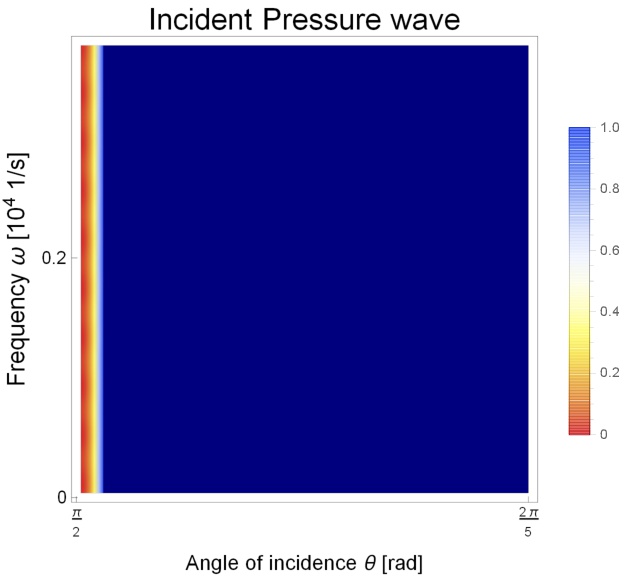}
		\caption{}	
	\end{subfigure}
	\hspace{1cm}
	\begin{subfigure}[H]{0.45\textwidth}
		\includegraphics[width=\textwidth]{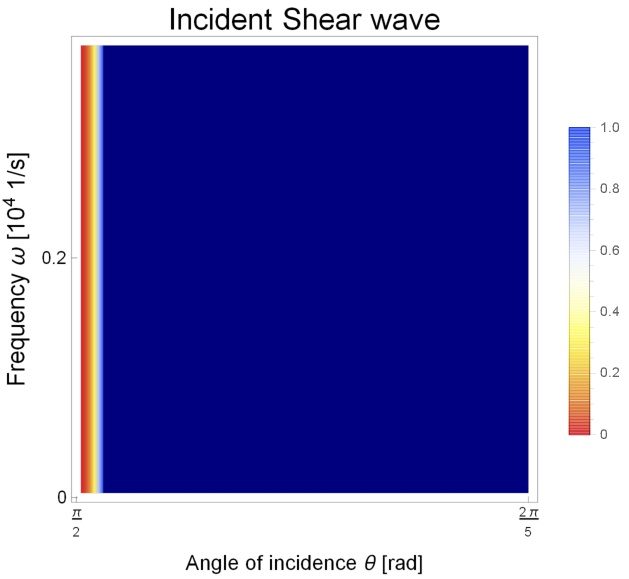}
		\caption{}
	\end{subfigure}
\caption{Analytical plot of a single interface separating two Cauchy materials \text{CM4} and \text{CM3} (no metamaterial's slab) reflection coefficient, as function of the angle of incidence and of the wave-frequency - (left): incident pressure wave; (right) incident shear wave.}
	\label{fig:sweep_20_cell_caso_b_no_slab}
\end{figure}

As a limit case, we also present in Fig.~\ref{fig:sweep_20_cell_caso_a_no_slab} and Fig.~\ref{fig:sweep_20_cell_caso_b_no_slab} the behaviour of a single interface separating two Cauchy materials \textit{CM3} and \textit{CM4} (no metamaterial's slab).

From Fig.~\ref{fig:sweep_20_cell_caso_a_no_slab} and Fig.~\ref{fig:sweep_20_cell_caso_b_no_slab}, we can once again retrieve the fact that the diode behaviour is effectively driven by the relative stiffness of the two homogeneous layers, while the embedded metamaterial slab has little effect on that.

However, except for this driving effect on the low-frequency diode behaviour, the structure's refractive response is completely different from that obtained in presence of the embedded metamaterial's slab.

By comparison with Fig.~\ref{fig:sweep_20_cell_caso_1} and Fig.~\ref{fig:sweep_20_cell_caso_2}, we instead remark that the presence of the metamaterial's slab strongly influence the medium-high frequency behaviour, transforming the diode in an acoustic screen that reflects waves independently of the direction of the incident wave.
This switch in the meta-structures reflective properties is clearly driven by the metamaterial's band-gap.

\begin{figure}[H]
	\centering
	\begin{subfigure}[H]{0.45\textwidth}
		\includegraphics[width=\textwidth]{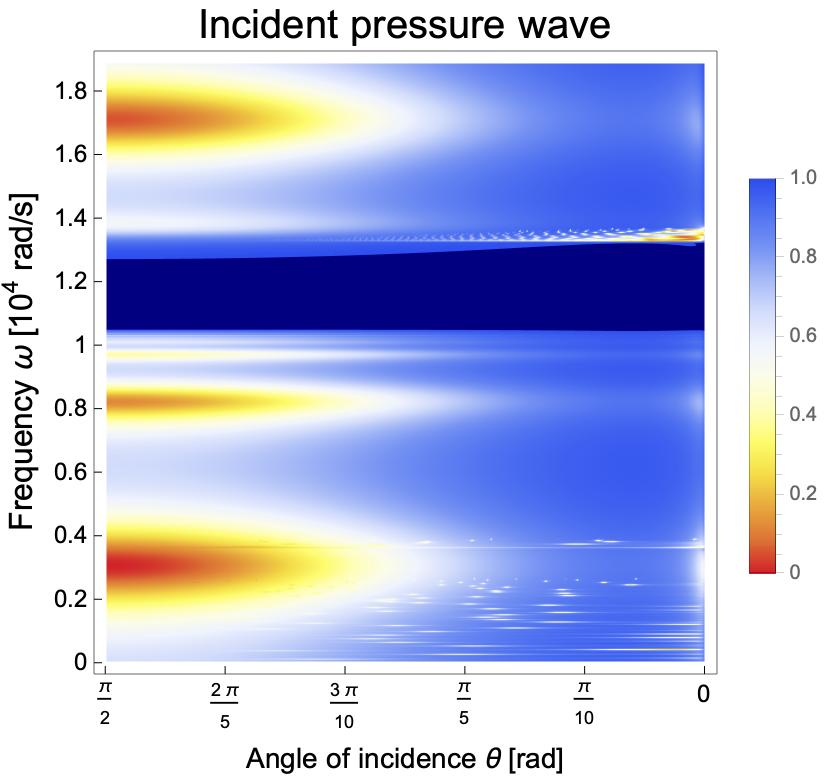}	
		\caption{}
	\end{subfigure}
	\hspace{1cm}
	\begin{subfigure}[H]{0.45\textwidth}
		\includegraphics[width=\textwidth]{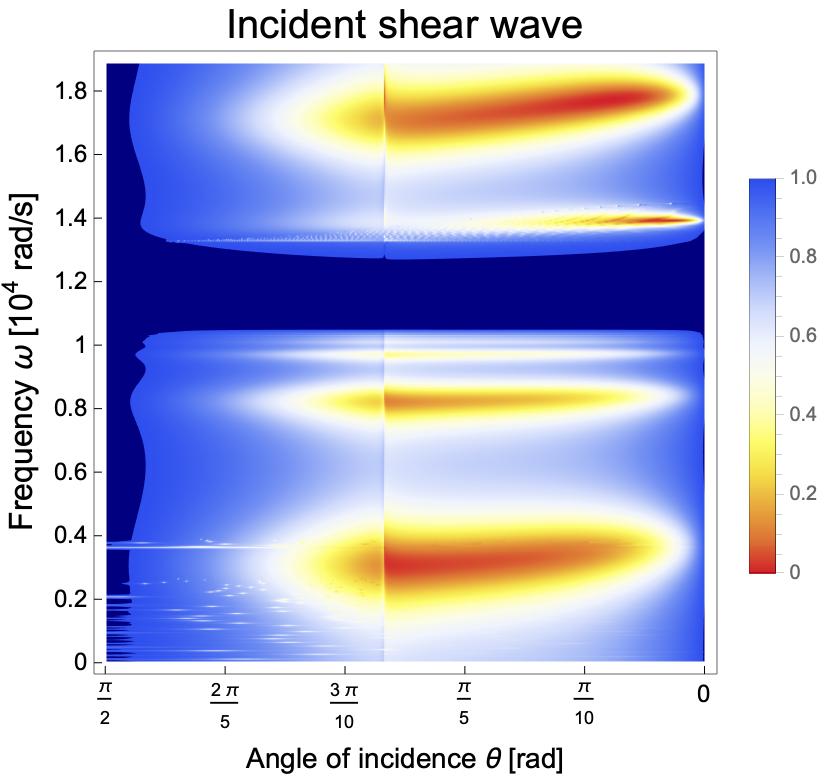}
		\caption{}
	\end{subfigure}
	\caption{Analytical plot of the micromorphic reflection coefficient for a 20 unit cells thick slab made up of \textit{MM1} material and embedded between the \textit{CM5} Cauchy (green in Fig.~\ref{fig:slab_stru_a}) and \textit{CM6} (blue in Fig.~\ref{fig:slab_stru_a}), as function of the angle of incidence and of the wave-frequency - (left): incident pressure wave; (right) incident shear wave.}
	\label{fig:sweep_20_cell_caso_1_poiss_posi}
\end{figure}

\begin{figure}[H]
	\centering
	\begin{subfigure}[H]{0.45\textwidth}
		\includegraphics[width=\textwidth]{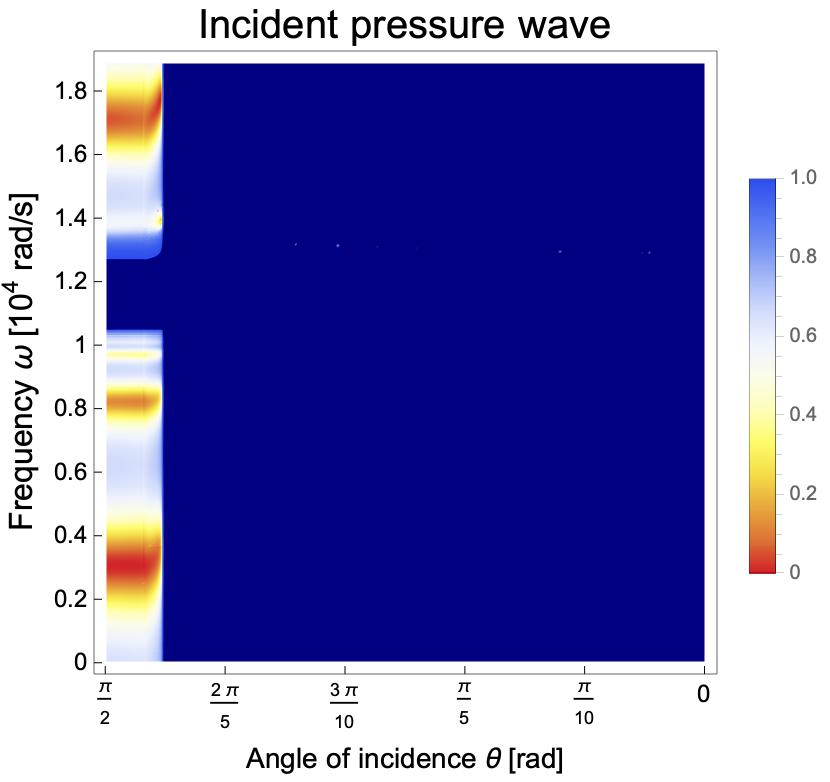}
		\caption{}	
	\end{subfigure}
	\hspace{1cm}
	\begin{subfigure}[H]{0.45\textwidth}
		\includegraphics[width=\textwidth]{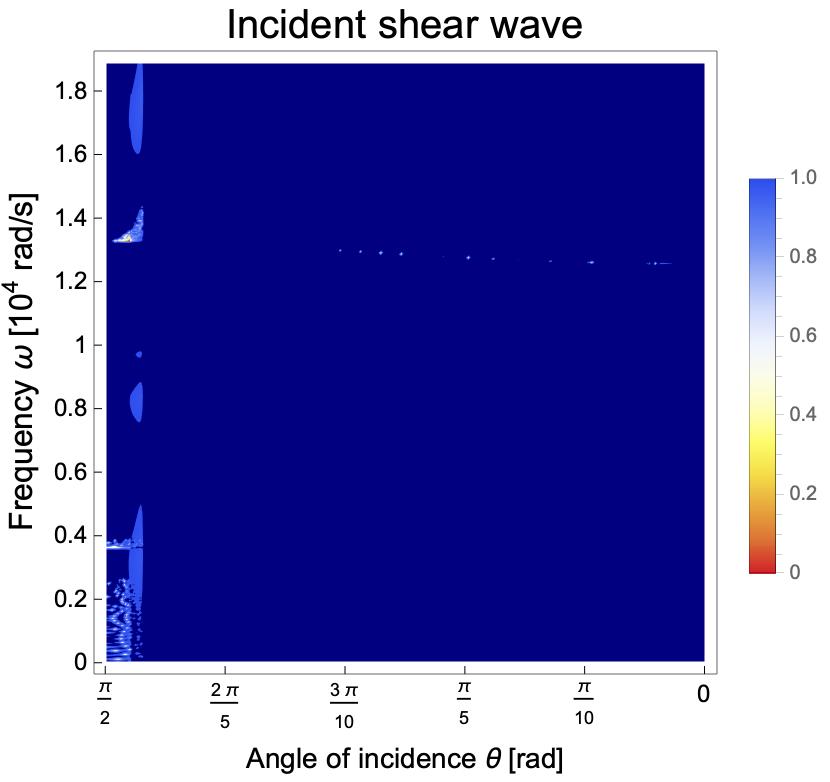}
		\caption{}
	\end{subfigure}
\caption{Analytical plot of the micromorphic reflection coefficient for a 20 unit cells thick slab made up of \textit{MM1} material and embedded between the \textit{CM6} Cauchy (green in Fig.~\ref{fig:slab_stru_b}) and \textit{CM5} (blue in Fig.~\ref{fig:slab_stru_b}), as function of the angle of incidence and of the wave-frequency - (left): incident pressure wave; (right) incident shear wave.}
	\label{fig:sweep_20_cell_caso_2_poiss_posi}
\end{figure}

We carry out one last example (see Fig.~\ref{fig:sweep_20_cell_caso_1_poiss_posi}-\ref{fig:sweep_20_cell_caso_2_poiss_posi}) in which the two isotropic Cauchy materials embedding the slab have a positive Poisson's ratio that can be calculated with the Lamé coefficients in Table~\ref{tab:parameters_Cau_ext_positi_Poisso}.
This last example helps to further clarify that the low-frequency diode behaviour is driven by the difference macro-stiffness of the two materials and is not related to the fact that the elastic coefficients are chosen to be close to singular values (see footnote 4).
\begin{table}[H]
	\renewcommand{\arraystretch}{1.5}
	\centering
		\begin{tabular}{|c|c|c|c|c|} 
			\hline
			$\mu_{\rm CM5}$ [Pa] & $\lambda_{\rm CM5}$ [Pa] & $\kappa_{\rm CM5}$ [Pa] & $\nu_{\rm CM5}$ [-] & $\rho_{\rm CM5}$ [kg/m$^3$]\\
			\hline
			$1.76\times10^{10}$ & $4.4\times10^{9}$ & $2.2\times10^{10}$ & $0.1$ & $4400$\\
			\hline
			\hline
			$\mu_{\rm CM6}$ [Pa] & $\lambda_{\rm CM6}$ [Pa] & $\kappa_{\rm CM6}$ [Pa] & $\nu_{\rm CM6}$ [-] & $\rho_{\rm CM6}$ [kg/m$^3$]\\
			\hline
			$1.76\times10^{8}$ & $4.4\times10^{7}$ & $2.2\times10^{8}$ & $0.1$ & $4400$\\
			\hline
		\end{tabular}
	\caption{Values of the density, the Lamé constants, Poisson's ratio and bulk modulus under the plane strain hypothesis for (\textit{top}) the isotropic Cauchy material \textit{CM5} and (\textit{bottom}) the isotropic Cauchy material \textit{CM6}.}
	\label{tab:parameters_Cau_ext_positi_Poisso}
\end{table}
Figure \ref{fig:sweep_20_cell_caso_1_poiss_posi} shows the reflective properties for the case in which the slab is embedded between two different Cauchy material \textit{CM5} Cauchy (green in Fig.~\ref{fig:slab_stru_a}) and \textit{CM6} (blue in Fig.~\ref{fig:slab_stru_a}), while figure \ref{fig:sweep_20_cell_caso_2_poiss_posi} shows the reflective properties for the specular case, in which the two Cauchy material are now switched.
\section{Conclusions}
In this paper we show the importance of disposing of a reduced model to enable the effective use of metamaterials in meta-structural design.
Indeed, the use of metamaterials for the conception of realistic structures is currently prevented by i) the computational impossibility of simulating the response of large-scale structures while coding all the microstructure's details and ii) the difficulty to deal with specimens of finite-size (well-posed boundary conditions often unavailable).
We show here that an enriched continuum model of the micromorphic type (Relaxed Micromorphic Model) can be effectively used to model metamaterials' response, even for specimens of finite-size. The reduced model's structure, coupled with the introduction of well-posed interface conditions allows us to unveil the response of meta-structures combining metamaterials and classical-materials bricks.
In particular, we are able to conceive a metamaterial/classical-material structure that acts as a mechanical diode for low/medium frequencies and as a total screen for higher frequencies. This could have, for example, important implication for the conception of large-scale structures that are protected from seismic waves coming from the exterior and that control vibrations coming from the interior.
While current studies mainly focus on the design of complex heterogeneous, asymmetric microstructures to obtain a mechanical diode, we show here that such a diode can be also obtained embedding metamaterials with symmetric microstructures between two homogeneous materials with different stiffness.
This paper lays the basis to widen our knowledge towards the conception of more and more complex, large-scale meta-structures that can control elastic waves and recover energy.
It is clear that to reach a refined quantitative prediction of metamaterials' response, the Relaxed Micromorphic model will need considerable extension to increase its precision up to very small wavenumbers and to enable a refined description of static and dynamic size effects.

\begingroup
\setstretch{1}
\setlength\bibitemsep{3pt}
\printbibliography
\endgroup

\end{document}